\documentclass[a4paper,10pt]{article}

\usepackage{amsmath}
\usepackage{amssymb}
\usepackage{amsthm}
\usepackage{amsfonts}
\usepackage{xcolor}
\usepackage{graphicx}
\usepackage{mathtools}
\usepackage{enumerate}
\usepackage{multirow}
\usepackage{rotating}
\usepackage{setspace}
\usepackage{lineno}
\usepackage{url}

\theoremstyle{theorem}
\newtheorem{thm}{Theorem}

\theoremstyle{definition}

\usepackage{caption}
\usepackage{booktabs}
\usepackage{subcaption}

\usepackage{geometry}

\begin{document}
	
	\title{\large{Optimal Vaccination and Treatment Strategies in Reduction of COVID-19 Burden }}
	
	\vspace{0.1in}
	\author{{\bf\large Bishal Chhetri$^{1,a}$,  D. K. K. Vamsi$^{*,a}$, S Balasubramanian$^{b}$, Carani B Sanjeevi$^{c, d}$}\hspace{2mm} \\
		{\it\small $^{a,b}$Department of Mathematics and Computer Science, Sri Sathya Sai Institute of Higher Learning, Prasanthi Nilayam}, \\
		{\it \small Puttaparthi, Anantapur District - 515134, Andhra Pradesh, India}\\
		{\it\small $^{c}$  Vice-Chancellor, Sri Sathya Sai Institute of Higher Learning -  SSSIHL, India}\\
		{\it\small $^{d}$ Department of Medicine, Karolinska Institute, Stockholm, Sweden }\\
		{\it\small bishalchhetri@sssihl.edu.in, dkkvamsi@sssihl.edu.in$^{*}$,}\\
		{\it\small sanjeevi.carani@sssihl.edu.in, sanjeevi.carani@ki.se}\\
		{\small $^{1}$  First Author},
		{ \small $^{*}$ Corresponding Author}
		\vspace{1mm}
	}
	
	\date{}
	\maketitle

\begin{abstract} \vspace{.25cm}

In this study, we formulate a mathematical model incorporating age specific transmission dynamics of COVID-19 to evaluate the role of vaccination and treatment strategies in reducing the size of  COVID-19 burden. Initially, we establish the positivity and boundedness of the solutions of the model and calculate the basic reproduction number. We  then formulate an optimal control problem with vaccination and  treatment as control variables. Optimal vaccination and treatment policies are analysed for different values of the weight constant associated with the cost of vaccination and different transmissibility levels. Findings from these suggested that the combined strategies(vaccination and treatment) worked best in minimizing the infection and disease induced mortality. In order to reduce COVID-19 infection and COVID-19 induced deaths to maximum, it was observed that optimal control strategy should be prioritized to population with age greater than 40 years. 
Not much difference was found between individual strategies and combined strategies in case of mild epidemic ($R_0 \in (0, 2)$). For higher values of $R_0  (R_0 \in (2, 10))$ the combined strategies was found to be best in terms of minimizing the overall infection. The infection curves  varying the efficacies of the vaccines were also analysed and it was found that higher  efficacy of the vaccine resulted in lesser number of infection and COVID induced death.

\end{abstract}
\section{Introduction}

 Mathematical modeling of infectious
diseases such as COVID-19, influenza, dengue, HIV/AIDS etc. is one of the most important research areas today. Mathematical epidemiology has contributed to a better
understanding of the dynamical behavior of these infectious diseases,
its impacts, and possible future predictions about its
spreading. Mathematical models are used in comparing,
planning, implementing, evaluating, and optimizing various
detection, prevention, therapy, and control programs. COVID-19 is one such contagious respiratory and vascular disease that has shaken the world today. It is  caused by severe acute respiratory syndrome coronavirus 2 (SARS-CoV-2). On 30 january it was declared as a Public Health Emergency of International Concern. As of latest statistics(on 24 January 2021) of COVDI-19, around 96.2 million cases has been reported and around 2 million have died worldwide. Several mathematical models has been developed  to understand the dynamics  of the disease. In \cite{chhetri2020within} a basic within host model is developed to determine the crucial inflammatory mediators and the role of combined drug therapy in the treatment of COVID-19. A SAIU compartmental mathematical model that explains the transmission dynamics of COVID-19 is developed in \cite{samui2020mathematical}.  The role of some of the control policies such as treatment, quarantine, isolation, screening, etc. are also applied to control the spread of infectious diseases \cite{kkdjou2020optimal,libotte2020determination,aronna2020model}. COVID-19 has caused the most severe health issues for adults over the age of 60 with particularly fatal results for those 80 years and older. This is due to the number of underlying health conditions present in older population \cite{ch}. A mathematical model for estimating the age-specific transmissibility of a novel coronavirus is developed in \cite{zhao2020mathematical}. In this study the age age-specific SEIARW model was fitted with the reported data well by dividing the population into four age groups and the results from this study suggested that the highest transmissibility occurred from age group $1-14$ to $15-44$.

One of the most effective method to prevent any infectious disease is vaccination. Implementation of vaccination program is estimated to prevent approximately 2-3 million deaths each year \cite{sari2017optimal}. With several stakeholders working together across the globe some of the countries are successful in producing COVID-19 vaccines today. The approved vaccines for COVID-19 today includes Pfizer with 95 $\%$ efficacy, Moderna with 94 $\%$ efficacy, and AstraZeneca-oxford with  70 $\%$ efficacy.  Drug Controller General of India (DCGI), the country’s national drug regulator, approved two coronavirus vaccines for restricted emergency use — Serum Institute of India’s Covishield (the Indian variant of the AZD1222 vaccine developed by Oxford University and AstraZeneca) and Bharat Biotech’s Covaxin \cite{nitisha}. Several mathematical models are developed to study the role of vaccination and treatments in reducing the disease burden. In \cite{bubar2020model} a mathematical model is used to compare five age-stratified prioritization strategies. A highly effective transmission-blocking vaccine prioritized to adults ages 20-49 years was found to minimize the cumulative incidence, whereas mortality and years of life lost were minimized in most scenarios when the vaccine was prioritized to adults over 60 years old.  Reports from Israel suggested that one dose of Pfizer vaccine could be less effective than expected \cite{mahase2021covid}. A two-dose regimen of BNT162b2 conferred 95$\%$ protection against Covid-19 in persons 16 years of age or older. Safety over a median of 2 months was similar to that of other viral vaccines  \cite{polack2020safety}. The mRNA-1273 vaccine showed 94.1$\%$ efficacy at preventing Covid-19 illness, including severe disease \cite{baden2020efficacy}. 
 The combination vaccines for protection against multiple diseases began with the combination of individual diphtheria, tetanus, and pertussis (DTP) vaccines into a single product; this combined vaccine was first to be used to vaccinate infants and children in 1948. Over the years we have seen the addition of other vaccines to the combination and the replacement of components to improve its reactogenicity profile \cite{skibinski2011combination}. The addition of inactivated polio, Haemophilus influenzae, and hepatitis B vaccines into the combination has facilitated the introduction of these vaccines into recommended immunization schedules by reducing the number of injections required and has therefore increased immunization compliance \cite{skibinski2011combination}.

 To reflect the real behavior of some infectious diseases and to make models more realistic, many researchers have proposed and analyzed more realistic models including delays to model different mechanisms in the dynamics of epidemics like latent
period, temporary immunity and length of infection \cite{khan2003epidemic,van1994some}. An optimal control problem
with time delay  in both the state variable and control variable is studied in \cite{elhia2013optimal}. 

Motivated by the above, in this study, we consider a nine compartment age structured model to study the  role of individual vaccines, combination vaccines  and treatment in reducing the COVID-19 infection. In the model we incorporate time delay in both the control and state variables.

The paper is organised as follows: In section 2 we formulate a mathematical model explaining the details of the parameters and variables used and establish the positivity and boundedness of the solutions. In section 3 we formulate an optimal control problem to evaluate the role  of vaccination and treatment in reducing the cumulative infection and disease induced mortality. Numerical simulation is presented in section 4 followed by discussion and conclusion in section  5.

\newpage

\section{Model Formulation}
Various mathematical models has been developed and studied to understand the dynamics of COVID-19 and design optimal control strategies to control an epidemic. In this work we  formulate an optimal  control problem with age specific transmission dynamics of COVID-19. The total population in the model is divided into different compartments such as susceptible$(S_i)$, vaccinated but not protected$(V_i)$, ineffectively vaccinated$(F_i)$, Protected$(P_i)$, exposed$(E_i)$, infected$(I_i)$, hospitalized$(J_i)$, recovered$(R_i)$ and deaths($D_i$) for $i=1,2$. We consider two age groups here, the first between 0-40 years and second group with age greater than 40 years.  At any point in time we assume that the individuals will be in one of these compartments. When susceptible individuals in age group $i$ come in  close contact with the infected or hospitalized they become exposed to the
virus at rates $\beta_{ij}$ where, $\beta_{ij}$ is
the transmission rate between age groups $i$ and $j$. Exposed individuals
$E_i$ progress to the infectious class $I_i$ at the rate k (where 1/k is the mean latent
period). The term $\alpha_i e^{-\gamma\tau_1}$ gives the rate at  which infected are hospitalized. Here $\tau_1$ represents the delay in hospitalization and with increasing value of delay or $\gamma$ the rate of movement to $J_i$ compartment is less \cite{maki2020delayed},  $d_{1i}$ and $\gamma$ are the disease induced death rate and recovery rate of the infected individuals. Hospitalized individuals either recover at the constant
rate $\gamma$ or die at the age-specific rate $d_{2i}$ .\\

We employ time-dependent (age-specific) control
functions to measure the effectiveness of age-specific vaccination and treatment policies aimed at
minimizing the number of infected individuals during the pandemic. The control functions $\mu_{1i}(t)$ and $\mu_{2i}(t)$ determine the age-specific vaccination rates
of susceptible individuals $(S_i ) $ per unit of time for each age group $i$. We assume that the suceptibles are given both the vaccines together at the same time and only those individuals who were vaccinated at time $(t-\tau)$ will now move to $V_i, F_i$ or $P_i$ compartment. The control variables $\mu_{3i}, \mu_{4i}$ represents the age specific treatment rates for infected and hospitalized population respectively. To make model realistic we assume that there is a time  lag between treatment and recovery represented by $\tau_2$ and $\tau_3$ for infected and hospitalized population respectively. The dynamic
model with age-specific controls is described by the following system of nonlinear
differential equations:

\begin{eqnarray}
\frac{dS_i}{dt}& =&\omega_i- \sum_{j=1}^{2} \beta_{ij}(I_j+J_j)S_i-\mu_{1i}(t-\tau)S_i(t-\tau)-\mu_{2i}(t-\tau)S_i(t-\tau)-\mu S_i\\
  \frac{dV_i}{dt}&=&\epsilon_{1i}\mu_{1i}(t-\tau)S_i(t-\tau)+\gamma_{1i}\mu_{2i}(t-\tau)S_i(t-\tau)-\sum_{j=1}^{2} \beta_{ij}(I_j+J_j)V_i -\mu V_i\\
  \frac{dF_i}{dt}&=&\epsilon_{2i}\mu_{1i}(t-\tau)S_i(t-\tau)+\gamma_{2i}\mu_{2i}(t-\tau)S_i(t-\tau)-\sum_{j=1}^{2} \beta_{ij}(I_j+J_j)F_i -\mu F_i\\
  \frac{dP_i}{dt}&=&(1-\epsilon_{1i}-\epsilon_{2i})\mu_{1i}(t-\tau)S_i(t-\tau)+(1-\gamma_{1i}-\gamma_{2i})\mu_{2i}(t-\tau)S_i(t-\tau)-\mu P_i\\
   \frac{dE_i}{dt}&=&\sum_{j=1}^{2} \beta_{ij}(I_j+J_j)\bigg(S_i+V_i+F_i\bigg)-k E_i-\mu E_i \\ 
   \frac{dI_i}{dt}&=&kE_i-d_{1i} I_i-\alpha_ie^{-\gamma \tau_1 }I_i(t-\tau_1)-\mu_{3i}(t-\tau_2)I_i(t-\tau_2) -\gamma I_i\\
  \frac{dJ_i}{dt}&=&\alpha_ie^{-\gamma \tau_1 }I_i(t-\tau_1)-d_{2i}J_i-\mu_{4i}(t-\tau_3)J_i(t-\tau_3)\\
   \frac{dR_i}{dt}&=&\gamma I_i+\mu_{4i}(t-\tau_3)J_i(t-\tau_3)+\mu_{3i}(t-\tau_2)I_i(t-\tau_2)-\mu R_i \\
   \frac{dD_i}{dt}&=&d_{1i}I_i+d_{2i}J_i -\mu D(i)
\end{eqnarray}

{\flushleft \bf{OBJECTIVES OF THE PROPOSED STUDY}} \vspace{.25cm}
\begin{itemize}
   
    \item [1.] 	To study and compare the dynamics of cumulative infection, hospitalized and mortality with and without  the controls. 
     \item [2.] To determine which age groups should be prioritized
for COVID pandemic vaccination.
       \item [3.] 	To study and compare the dynamics of infected and   hospitalized population with varying efficacies of the vaccine.
       \item [4.] 	To study and compare the dynamics of infected and   death population with varying cost of implementation of vaccination strategy.
 
\end{itemize} \vspace{.25cm}

\newpage

 \begin{table}[ht!]
     	\caption{}
     	\centering 
     	\begin{tabular}{|l|l|} 
     		\hline\hline
     		
     		\textbf{Symbols} &  \textbf{Biological Meaning} \\  
     		\hline\hline 
     		$S_i$ & Suceptible Population\\
     		\hline\hline
     		
     		$V_i$ & effectively Vaccinated but not protected \\
     		\hline\hline
     		$F_i$ & ineffectively vaccinated \\
     		\hline\hline
     		$P_i$ & Protected Population \\
     		\hline\hline
     		$E_i$ & Exposed Population  \\
     			\hline\hline
     		$I_i$ & infected Population  \\
     			\hline\hline
     		$J_i$ & hospitalized Population\\
     			\hline\hline
     		$R_i$ & recovered Population  \\
     			\hline\hline
     		$\omega_i$ & Rate of entries in each groups  \\
     		\hline\hline
     		$\beta_{ij}$ & transmission rates among different age groups \\
     		\hline\hline
     		$\mu_{1i}$ & rate of decrease in suceptibles due to first vaccine \\
     			\hline\hline
     		$\mu_{2i}$ & rate of decrease in suceptibles due to second vaccine \\
     		\hline\hline
     		$\mu$ & Natural death rate  \\
     		\hline\hline
     		$d_{11} $ & disease induced death rates for first infected population  \\
     		\hline\hline
     		$d_{12}$ & death for second infected group \\
     		\hline\hline
     		$d_{21}$ & disease induced death rates for first group hospitailized population  \\
     		\hline\hline
     		$d_{22}$& disease induced death rates for second group hospitailized population \\
     			\hline\hline
     		$k$& infection rates \\
     			\hline\hline
     		$\alpha_i$& rates at which infected are hospitalized  \\
     			\hline\hline
     		
     		$\mu_{3i}$& recovery rate of infected due to treatment \\
     			\hline\hline
     		$\mu_{4i}$&recovery rate of hospitalized due to treatment   \\ 
     		\hline\hline
     		$\epsilon_{1i}, \epsilon_{2i}$&efficacy of first vaccine   \\ 
     		\hline\hline
     		$\gamma_{1i}, \gamma_{2i}$& efficacy of second vaccine  \\ 
     		
     		\hline\hline
     		$\gamma$& natural recovery rate  \\
     		
     	\hline\hline
     		
     \end{tabular}
     \end{table} \vspace{.25cm}
     
     \newpage
	\textbf{Positivity and Boundedness}\\
For any mathematical model it is fundamental to show that the system of equations considered are positive and has bounded solutions. We now show that if the initial conditions of the system (3.1)-(3.9) are positive, then the solution remain positive for any future time. Using the  equations (3.1)-(3.9), we get,
\begin{align*}
\frac{dS_i}{dt} \bigg|_{S_i=0} &\geq 0 ,  &  
\frac{dV_i}{dt} \bigg|_{V_i=0} &= \epsilon_{1i}\mu_{1i}S_i +\gamma_{1i}\mu_{2i}S_i \geq 0,
\\ \\
\frac{dF_i}{dt} \bigg|_{F_i=0} &= \epsilon_{2i}\mu_{1i}S_i +\gamma_{2i}\mu_{2i}S_i  \geq 0.  &
\frac{dD_i}{dt}&= d_{1i}I_i+d_{2i}J_i\bigg|_{J_i=0}\geq 0 \\\\
\frac{dE_i}{dt} \bigg|_{E_i=0} &=  \sum_{j=1}^{2} \beta_{ij}(I_j+J_j)\bigg(S_i+V_i+F_i\bigg) \geq 0.  &
\frac{dI_i}{dt} \bigg|_{I_i=0} &= k E_{i}  \geq 0.  & \\
\\
\frac{dJ_i}{dt} \bigg|_{J_i=0} &= \alpha_i e^{-\gamma\tau_1} I_{i}  \geq 0.  &
\frac{dR_i}{dt} \bigg|_{R_i=0} &= \gamma I_i+\mu_{4i}J_i+\mu_{3i}I_i  \geq 0.  & \\\\
\frac{dP_i}{dt} \bigg|_{P_i=0} &= (1- \epsilon_{1i}-\epsilon_{2i})\mu_{1i}S_i +(1-\gamma_{1i}-\gamma_{2i})\mu_{2i}S_i  \geq 0.  &
\end{align*}

\vspace{1.5mm}
\noindent
\\ 
Thus all the above rates are non-negative on the bounding planes (given by $S_i=0$, $V_i=0$, $P_i=0$,$F_i=0$,$E_i=0$,$I_i=0$,$J_i=0$,$R_i=0$,$D_i=0$) of the non-negative region of the real space. So, if a solution begins in the interior of this region, it will remain inside it throughout time $t$. This  happens because the direction of the vector field is always in the inward direction on the bounding planes as indicated by the above inequalities. Hence, we conclude that all the solutions of the the system (3.1)-(3.9) remain positive for any time $t>0$  provided that the initial conditions are positive. Next we will show that the solution is bounded with each of the bounded control variables.  \vspace{.25cm}

\underline{\textbf{Boundedness}}\textbf{:}
Let  $N_i(t) = S_i(t)+V_i(t)+F_i(t)+P_i+E_i+I_i+J_i+R_i+ D_i$ \\

Now,  
\begin{equation*}
\begin{split}
\frac{dN_i}{dt} & = \frac{dS_i}{dt} +  \frac{dV_i}{dt}+  \frac{dF_i}{dt} +  \frac{dP_i}{dt} +  \frac{dE_i}{dt}+  \frac{dI_i}{dt}+ \frac{dJ_i}{dt} +  \frac{dR_i}{dt}+  \frac{dD_i}{dt} \\[4pt]
& = \bigg(\omega_i+\mu(I_i+J_i)\bigg)-\mu N(t)\\
& \leq 0
\end{split}
\end{equation*} with the assumption that $\bigg(\omega_i+\mu(I_i+J_i)\bigg) \leq \mu N(t)$.
This implies that $N_i(t)=C$, where C is a constant

 Thus we have shown that the system (2.1)-(2.9) is positive and bounded for each bounded controls considerd. Therefore the biologically feasible region is given by the following set, 
\begin{equation*}
\Omega = \bigg\{\bigg(S_i(t), V_i(t), P_i(t),F_i(t), E_i(t),I_i(t),J_i(t),R_i(t),D_i(t)\bigg)  : N_i(t) \leq C, \ t \geq 0 \bigg\}
\end{equation*}

\subsection{\textbf{Calculation of Basic Reproduction Number $R_0$}}
     
     The basic reproduction number which is the average number of secondary cases produced per primary case is calculated using the next generation matrix method described in \cite{diekmann2010construction} at infection free equilibrium. Our system (2.1)-(2.9) has four infected states $(E_1, E_2,I_1,I_2)$. In order to see the behaviour of the optimal strategies with varying transmissibility we calculate the basic reproduction number.   Calculating the jacobian matrix at infection free equilibrium $E_0$(which has only susceptible component) we have,
    
\begin{equation*}
J(E_0)=    
\begin{bmatrix}
-k-\mu & 0 & \beta_{11} S_1^* & \beta_{12} S_1^*  \\[6pt]
0 & -k-\mu & \beta_{11}S_2^*& \beta_{12} S_2^*  \\[6pt]
k& 0&  -d_{11}-\gamma-\alpha_1e^{-\gamma\tau_1}& 0\\[6pt]
0 & k & 0 & -d_{12}-\gamma-\alpha_2 e^{-\gamma\tau_1} \\[6pt]
\end{bmatrix}
\end{equation*}      
or,
$$J(E_0) = F + V$$ where, $F$ describes transmission of new infection and $V$ describes changes in the state including removal by death or recovery rate.

Matrix $F$ and $V$ are given as,\\
\begin{equation*}
F =    
\begin{bmatrix}
0 & 0 & \beta_{11} S_1^* & \beta_{12} S_1^*  \\[6pt]
0 & 0 & \beta_{11}S_2^*& \beta_{12} S_2^*  \\[6pt]
0& 0& 0 & 0 \\[6pt]
0& 0& 0 & 0\\[6pt]
\end{bmatrix}
\end{equation*}

\begin{equation*}
V =   
\begin{bmatrix}
-k-\mu & 0 & 0& 0  \\[6pt]
0 & -k-\mu & 0& 0  \\[6pt]
k& 0&  -d_{11}-\gamma-\alpha_1e^{-\gamma\tau_1}& 0\\[6pt]
0 & k & 0 & -d_{12}-\gamma-\alpha_2 e^{-\gamma\tau_1} \\[6pt]
\end{bmatrix} 
\end{equation*}

Calculating the inverse of $V$ we get,
\begin{equation*}
V^{-1} =   
\begin{bmatrix}
\frac{1}{-k-\mu} & 0 & 0& 0  \\[6pt]
0 & \frac{1}{ -k-\mu} & 0& 0  \\[6pt]
\frac{-k}{(k+\mu)(d_{11}+\gamma+\alpha_1 e^{-\gamma \tau_1})}& 0&  \frac{1}{-d_{11}-\gamma-\alpha_1e^{-\gamma\tau_1}}& 0\\[6pt]
0 & \frac{-k}{(k+\mu)(d_{11}+\gamma+\alpha_1 e^{-\gamma \tau_1})} & 0 & -d_{12}-\gamma-\alpha_2 e^{-\gamma\tau_1} \\[6pt]
\end{bmatrix} 
\end{equation*}
Now

\begin{equation*}
-FV^{-1} =    
\begin{bmatrix}
\frac{\beta_{11}k S_1^*}{p} & \frac{\beta_{12}k S_1^*}{q} & \frac{\beta_{11} S_1^*}{p}&\frac{\beta_{12} S_1^*}{(k+\mu)(q}\\[6pt]
\frac{\beta_{11}k S_2^*}{p} & \frac{\beta_{12}k S_2^*}{q} & \frac{\beta_{11} S_2^*}{p}&\frac{\beta_{12} S_2^*}{q} \\[6pt]
0 & 0 & 0 & 0\\[6pt]
0 & 0 & 0 & 0\\[6pt]
\end{bmatrix}
\end{equation*} 
where $$p=(k+\mu)(d_{11}+\gamma+\alpha_1 e^{-\gamma\tau_1})$$
$$q=(k+\mu)(d_{12}+\gamma+\alpha_2 e^{-\gamma\tau_1})$$
Since the last two rows of matrix $-FV^{-1}$ has all zeros as discussed in \cite{diekmann2010construction} we define an auxillary matrix and new matrix $K$ as,

\begin{equation*}
E =    
\begin{bmatrix}
1 & 0 \\[6pt]
0 & 1 & \\[6pt]
0 & 0 \\[6pt]
0 & 0 & \\[6pt]
\end{bmatrix}
\end{equation*}
 
 \begin{equation*}
K =    
\begin{bmatrix}
\frac{\beta_{11}k S_1^*}{p} & \frac{\beta_{12}k S_1^*}{q} \\[6pt]
\frac{\beta_{11}k S_2^*}{p} & \frac{\beta_{12}k S_2^*}{q}  \\[6pt]
\end{bmatrix}
\end{equation*} 
   
   Therefore the basic reproduction number which is defined as the spectral radius of $K$ is given by,\\
     
     \begin{equation*}
         \mathbf{ R_{0}}= \frac{\beta_{11}k S_1^*}{(k+\mu)(d_{11}+\gamma+\alpha_1 e^{-\gamma\tau_1})} + \frac{\beta_{12}k S_2^*}{(k+\mu)(d_{12}+\gamma+\alpha_2 e^{-\gamma\tau_1})}
     \end{equation*}

\section{Optimal Control Problem}
 
 Now we frame an optimal control problem with vaccination and treatment as controls. Our aim is to study the role and efficacies of these controls and design an optimal control policy that minimizes that infection and disease caused mortality. The controls that we consider are as follows:\\
 
 1. \textbf{Vaccination:} Vaccination is the most effective method of preventing infectious diseases. The susceptible sub population are given vaccine  to stimulates the body's immune system to recognize the agent as a threat and destroy it, thereby preventing transmission of
the disease among susceptible individual. Vaccination  also further helps in recognizing and destroying any of the microorganisms associated with that agent that it may encounter in the future. The first control that we consider here is vaccination. We assume that combination of vaccines is given to an infected individual and denote it by variable $\mu_{1i}$ (first vaccine) and $\mu_{2i}$ (second vaccine) for two age groups respectively. \\

2. \textbf{Treatment:} Infected and Hospitalized sub-population are given treatment
to reduce the burden of disease and control the spread of infection. Studies  in \cite{chhetri2020within}  suggested the combined use of immunomodulators and antiviral agents as a best treatment strategy to reduce the burden of COVID-19. Therefore the second control that we consider here is treatments to infected and hospitalized population. These treatments could be either immunomodelators such as INF, to boost the immune response or anti viral agents like remdesivir, arbidol etc. that inhibits the viral replication. We  denote this control variable by $\mu_{3i}$ and $\mu_{4i}$.

Let $U_1=(\mu_{11},\mu_{12})$, $U_2=(\mu_{21}, \mu_{22})$, $U_3=(\mu_{31}, \mu_{32})$ and $U_4=(\mu_{41},\mu_{42})$\\

The set of all admissible controls is given by \\
		$U = \left\{(U_1,U_2,U_3, U_4) : U_1 \in [0,U_1 max] , U_2 \in [0,U_2 max] , U_3 \in [0,U_3 max], U_4\in [0,U_4 max],t \in [0,T] \right\}$
	
	In order to reduce the complexity of the problem here we choose to model the control efforts via a linear combination of the quadratic terms. Also when the objective function is quadratic with respect to the control, diﬀerential equations arising from optimization have a known solution. Other functional forms sometimes lead to systems of diﬀerential equations that are diﬃcult to solve (\cite{djidjou2020optimal}, \cite{lee2010optimal}). Based on these we now propose and  define the optimal control problem with the goal to reduce the cost functional defines as follows,

	$	J(U_1, U_2, U_3, U_4) = \int_{0}^{T} \bigg(I_1(t)+I_2(t) + A_{1}(\mu_{11}(t)^2+\mu_{12}(t)^2)+A_{2}(\mu_{21}(t)^2+\mu_{22}(t)^2) \\
	\hspace{2cm}+A_{3}(\mu_{31}(t)^2+\mu_{32}(t)^2)	+A_4(\mu_{41}(t)^2+\mu_{42}(t)^2)\bigg) dt  \hspace{4cm} (3)$ 
	
	such that $u = \bigg(\mu_{11}(t),\mu_{12}(t),\mu_{21}(t),\mu_{22}(t),\mu_{31}(t),\mu_{32}(t),\mu_{41}(t),\mu_{42}(t)\bigg) \in U$ \\
	
	subject to the system 
	
	\begin{eqnarray}
\frac{dS_i}{dt}& =&\omega_i - \sum_{j=1}^{2} \beta_{ij}(I_j+J_j)S_i-\mu_{1i}(t-\tau)S_i(t-\tau)-\mu_{2i}(t-\tau)S_i(t-\tau)-\mu S_i\\
  \frac{dV_i}{dt}&=&\epsilon_{1i}\mu_{1i}(t-\tau)S_i(t-\tau)+\gamma_{1i}\mu_{2i}(t-\tau)S_i(t-\tau)-\sum_{j=1}^{2} \beta_{ij}(I_j+J_j)V_i-\mu V_i \\
  \frac{dF_i}{dt}&=&\epsilon_{2i}\mu_{1i}(t-\tau)S_i(t-\tau)+\gamma_{2i}\mu_{2i}(t-\tau)S_i(t-\tau)-\sum_{j=1}^{2} \beta_{ij}(I_j+J_j)F_i-\mu F_i \\
  \frac{dP_i}{dt}&=&(1-\epsilon_{1i}-\epsilon_{2i})\mu_{1i}(t-\tau)S_i(t-\tau)+(1-\gamma_{1i}-\gamma_{2i})\mu_{2i}(t-\tau)S_i(t-\tau)-\mu P_i \\
   \frac{dE_i}{dt}&=&\sum_{j=1}^{2} \beta_{ij}(I_j+J_j)\bigg(S_i+V_i+F_i\bigg)-k E_i-\mu E_i \\ 
   \frac{dI_i}{dt}&=&kE_i-d_{1i} I_i-\alpha_ie^{-\gamma \tau_1 }I_i(t-\tau_1)-\mu_{3i}(t-\tau_2)I_i(t-\tau_2) -\gamma I_i\\
  \frac{dJ_i}{dt}&=&\alpha_ie^{-\gamma \tau_1 }I_i(t-\tau_1)-d_{2i}J_i-\mu_{4i}(t-\tau_3)J_i(t-\tau_3)\\
   \frac{dR_i}{dt}&=&\gamma I_i+\mu_{4i}(t-\tau_3)J_i(t-\tau_3)+\mu_{3i}(t-\tau_2)I_i(t-\tau_2)-\mu R_i \\
   \frac{dD_i}{dt}&=&d_{1i}I_i+d_{2i}J_i -\mu D_i
\end{eqnarray}

	Here, the cost function (3) represents the number of total infected cells, and the overall cost for the implementation vaccines and treatments. Effectively, our aim is to minimize the total infected population and the overall cost.
	The integrand of the cost function (3), denoted by 
	$L(S,I,V,U_1,U_2,U_3) = 
	\bigg(I_1(t)+I_2(t) + A_{1}(\mu_{11}(t)^2+\mu_{12}(t)^2)+A_{2}(\mu_{21}(t)^2+\mu_{22}(t)^2) +A_{3}(\mu_{31}(t)^2+\mu_{32}(t)^2)	+A_4(\mu_{41}(t)^2+\mu_{42}(t)^2)\bigg)$
		is called the Lagrangian or the running cost.
	
 	The admissible solution set for the Optimal Control Problem (3)-(3.9) is given by
	$$\Omega = \{ (S_i, V_i, F_i, P_i, E_i,I_i,J_i,R_i,E_i,  U_1,U_2,U_3,U_4)\;| S_i, V_i, F_i, P_i, E_i, I_i, J_i, R_i, D_i \;   \text{ satisfy} (3.1)-(3.9) \} $$ 
	
	for all $u \in U$

	

	{\textbf{EXISTENCE OF OPTIMAL CONTROL}}\vspace{.25cm}\\
	
		We will show the existence of optimal control functions that minimize the cost functions within a finite time span $[0,T]$ showing that we satisfy the conditions stated in Theorem 4.1 of \cite{Wendell}.

	\begin{thm}
		There exists a 8-tuple of optimal controls $\bigg(\mu_{11}^{*}(t) , \mu_{12}^{*}(t) , \mu_{21}^{*}(t),\mu_{22}^{*}(t),\mu_{31}^{*}(t),\mu_{32}^{*}(t),\\
		\mu_{41}^{*}(t),\mu_{42}^{*}(t)\bigg)$ in the set of admissible controls U such that the cost functional is minimized i.e., 
		
		$$J[U_1^*, U_2^*,U_3^*, U_4^*] = \min_{(U_1^*,U_2^*,U_3^*, U_4^*) \in U} \bigg \{ J[U_1^*,U_2^*,U_3^*, U_4^*]\bigg\}$$ 
		corresponding to the optimal control problem (3)-(3.9).
	\end{thm}

	\begin{proof}
		
		 In order to show the existence of optimal control functions, we will show that the following conditions are satisfied : 
		
		\begin{enumerate}
			\item  The solution set for the system (3.1)-(3.9) along with bounded controls must be non-empty, $i.e.$, $\Omega \neq \phi$.
			
			\item  U is closed and convex and system should be expressed linearly in terms of the control variables with coefficients that are functions of time and state variables.
			
			\item The Lagrangian L should be convex on U and $L(S_i,V_i,F_i,P_i,E_i,I_i,J_i,R_i,D_i) \geq g(U_1,U_2,U_3,U_4)$, where $ g(U_1,U_2,U_3,U_4)$ is a continuous function of control variables such that $| (U_1,U_2,U_3,U_4)|^{-1}$ $ g(U_1,U_2,U_3,U_4)\to \infty$ whenever  $| (U_1,U_2,U_3,U_4)| \to \infty$, where $|.|$ is an $l^2(0,T)$ norm.
		\end{enumerate}

		Now we will show that each of the conditions are satisfied : 
		
		1. From Positivity and boundedness of solutions of the system (3.1)-(3.9), all solutions are bounded for each bounded control variable in $U$. Also clearly the RHS of the system (3.1)-(3.9) is lipschitz continuous. Using Picard-Lindelof Theorem\cite{makarov2013picard}, we have satisfied condition 1.

			2. $U$ is closed and convex by definition. Also, the system (3.1)-(3.9) is clearly linear with respect to controls such that coefficients are only state variables or functions dependent on time. Hence condition 2 is satisfied.
		
		3. Choosing $g(U_1,U_2,U_3,U_4)$ = $c(\mu_{11}^2+\mu_{12}^2+\mu_{21}^2+\mu_{22}^2+\mu_{31}^2+\mu_{32}^2+\mu_{41}^2+\mu_{42}^2)$ such that $c = min\left\{A_{1},A_{2},A_{3},A_4\right\}$, we can satisfy the condition 3.
		
		Hence there exists a control 8-tuple $(\mu_{11}^2+\mu_{12}^2+\mu_{21}^2+\mu_{22}^2+\mu_{31}^2+\mu_{32}^2+\mu_{41}^2+\mu_{42}^2))\in U$ that minimizes the cost function (3).
	\end{proof}

	\textbf{CHARACTERIZATION OF OPTIMAL CONTROL}\vspace{.25cm}
   
  We will obtain the necessary conditions for optimal control functions using the Pontryagin's Maximum Principle with delay in state and control variables \cite{gollmann2009optimal} and also obtain the characteristics of the optimal controls.
	
	The Hamiltonian for this problem is given by 
	
\begin{equation*}
\begin{split}
H & = \sum_{j=1}^{2}\bigg(I_i+A_1(\mu_{1i}^2(t)+A_2\mu_{2i}^2(t)+A_3\mu_{3i}^2(t)+A_4\mu_{4i}^2(t))\bigg) \\[4pt]
& + \sum_{j=1}^{2}\lambda_{S_i} \frac{dS_i}{dt}+ \sum_{j=1}^{2}\lambda_{V_i} \frac{dV_i}{dt}+ \sum_{j=1}^{2}\lambda_{F_i} \frac{dF_i}{dt}+ \sum_{j=1}^{2}\lambda_{P_i} \frac{dP_i}{dt}+ \sum_{j=1}^{2}\lambda_{E_i} \frac{dE_i}{dt}+ \sum_{j=1}^{2}\lambda_{I_i} \frac{dI_i}{dt}\\
& +  \sum_{j=1}^{2}\lambda_{J_i} \frac{dJ_i}{dt}+ \sum_{j=1}^{2}\lambda_{R_i} \frac{dR_i}{dt}\\
\end{split}
\end{equation*}
	
	Here $\lambda$ = ($\lambda_{S_i}$,$\lambda_{V_i}$,$\lambda_{F_i}$,$\lambda_{P_i}$,$\lambda_{E_i}$,$\lambda_{I_i}$,$\lambda_{J_i}$,$\lambda_{R_i}$) is called co-state vector or adjoint vector.
	
	Now the Canonical equations that relate the state variables to the co-state variables are  given by 
	
	\begin{equation}
	\begin{aligned}
	 \frac{\mathrm{d} \lambda _{S_i}}{\mathrm{d} t} &= -\frac{\partial H}{\partial S_i}-\chi_{[0,T-\tau]}(t)\frac{\partial H(t+\tau)}{\partial S_i(t-\tau)}\\
	 \frac{\mathrm{d} \lambda _{V_i}}{\mathrm{d} t} &= -\frac{\partial  H}{\partial{V_i} } \\
	  \frac{\mathrm{d} \lambda _{F_i}}{\mathrm{d} t} &= -\frac{\partial  H}{\partial{F_i} } \\
	   \frac{\mathrm{d} \lambda _{P_i}}{\mathrm{d} t} &= -\frac{\partial  H}{\partial{P_i} } \\
	    \frac{\mathrm{d} \lambda _{E_i}}{\mathrm{d} t} &= -\frac{\partial  H}{\partial{E_i} } \\
	     \frac{\mathrm{d} \lambda _{I_i}}{\mathrm{d} t} &= -\frac{\partial H}{\partial I_i}-\chi_{[0,T-\tau_1]}(t)\frac{\partial H(t+\tau_1)}{\partial I_i(t-\tau_1)}-\chi_{[0,T-\tau_2]}(t)\frac{\partial H(t+\tau_2)}{\partial I_i(t-\tau_2)}\\
	      \frac{\mathrm{d} \lambda _{J_i}}{\mathrm{d} t} &= -\frac{\partial H}{\partial J_i}-\chi_{[0,T-\tau_3]}(t)\frac{\partial H(t+\tau_3)}{\partial J_i(t-\tau_3)}\\
	           \frac{\mathrm{d} \lambda _{R_i}}{\mathrm{d} t} &= -\frac{\partial H}{\partial R_i}\\
	      \end{aligned}
	\end{equation}

	
	Substituting the Hamiltonian value gives the canonical system 
	
	\begin{equation*}
\begin{aligned}
	\frac{\mathrm{d} \lambda _{S_i}}{\mathrm{d} t} &= \bigg(\sum_{j=1}^{2}\beta_{ij}(I_j+J_j)+\mu\bigg)\lambda _{S_i}-\chi_{[0,T-\tau]}(t)(-\mu_{1i}-\mu_{2i})\lambda_{S_i}(t+\tau)\\
	&-\chi_{[0,T-\tau]}(t)(\epsilon_{1i}\mu_{1i}+\gamma_{1i}\mu_{2i})\lambda_{V_i}(t+\tau)-\chi_{[0,T-\tau]}(t)(\epsilon_{2i}\mu_{1i}+\gamma_{2i}\mu_{2i})\lambda_{F_i}(t+\tau)\\
	&-\chi_{[0,T-\tau]}(t)(1-\epsilon_{1i}-\epsilon_{2i})\mu_{1i}+(1-\gamma_{1i}-\gamma_{2i})\mu_{2i})\lambda_{P_i}(t+\tau)-\sum_{j=1}^{2}\beta_{ij}(I_j+J_j)\lambda_{E_i}\\
	\frac{\mathrm{d} \lambda _{V_i}}{\mathrm{d} t} &= \bigg(\sum_{j=1}^{2}\beta_{ij}(I_j+J_j)+\mu\bigg)\lambda _{V_i}-\bigg(\sum_{j=1}^{2}\beta_{ij}(I_j+J_j)\bigg)\lambda _{E_i}\\
	\frac{\mathrm{d} \lambda _{F_i}}{\mathrm{d} t} &= \bigg(\sum_{j=1}^{2}\beta_{ij}(I_j+J_j)+\mu\bigg)\lambda _{F_i}-\bigg(\sum_{j=1}^{2}\beta_{ij}(I_j+J_j)\bigg)\lambda _{E_i}\\
	\frac{\mathrm{d} \lambda _{P_i}}{\mathrm{d} t} &=-\mu \lambda_{P_i}\\
	\frac{\mathrm{d} \lambda _{E_i}}{\mathrm{d} t} &=(k+\mu) \lambda_{E_i}-k\lambda_{I_i}\\
	\frac{\mathrm{d} \lambda _{I_i}}{\mathrm{d} t} &=-1+(d_{1i}+\gamma)\lambda_{I_i}-\gamma \lambda_{R_i}+ \bigg(\sum_{j=1}^{2}\beta_{ij}S_j(\lambda _{S_j}-\lambda_{E_j})\bigg)
	+\bigg(\sum_{j=1}^{2}\beta_{ij}V_j(\lambda _{V_j}-\lambda_{E_j})\bigg)\\
	&+\bigg(\sum_{j=1}^{2}\beta_{ij}F_j(\lambda _{F_j}-\lambda_{E_j})\bigg)
	+\chi_{[0,T-\tau_1]}(t)\bigg(\alpha_i e^{-\gamma \tau_1} \lambda_{I_{i}}(t+\tau_1)-\alpha_i e^{-\gamma\tau_1}\lambda_{J_{i}}\bigg)\\
	&+\chi_{[0,T-\tau_2]}(t)\bigg( \mu_{3i}(\lambda_{I_{i}}(t+\tau_2)-\lambda_{R_{i}}(t+\tau_2))\bigg)\\
	\frac{\mathrm{d} \lambda _{J_{i}}}{\mathrm{d} t} &=(d_{2i})\lambda_{J_{i}}+ (\sum_{j=1}^{2}\beta_{ij}\bigg(S_j(\lambda _{S_{j}}-\lambda_{E_{j}})
	+V_j(\lambda _{V_{j}}-\lambda_{E_{j}})
	+F_j(\lambda _{F_j}-\lambda_{E_j})\bigg)\\
&+\chi_{[0,T-\tau_3]}(t)\bigg( \mu_{4i}(\lambda_{J_{i}}(t+\tau_3)-\lambda_{R_{i}}(t+\tau_3))\bigg)\\
\frac{\mathrm{d} \lambda _{R_i}}{\mathrm{d} t} &=-\mu \lambda_{R_i}\\
\end{aligned}
\end{equation*}

	along with transversality conditions
	$ \lambda _{S_i} (T) = 0, \  \lambda _{V_i} (T) = 0, \  \lambda _{F_i} (T) = 0,  \lambda _{P_i} (T) = 0, \  \lambda _{E_i} (T) = 0, \  \lambda _{I_i} (T) = 0,  \  \lambda _{J_i} (T) = 0, \  \lambda _{R_i} (T) = 0.$
	
	Now, to obtain the optimal controls, we will use the Hamiltonian minimization condition. Differentiating the Hamiltonian with respect to each of the controls and solving the equations, we obtain the optimal controls in the following. 
	Let $$x_i=(1-\epsilon_{1i}-\epsilon_{2i})S_1\lambda_{P_1}(t+\tau),i=1,2$$ 
	$$y_i=(1-\gamma_{1i}-\gamma_{2i})\lambda_{P_i}(t+\tau),i=1,2$$
	
	\begin{eqnarray*}
	\mu_{11}^{*} &=& \min\bigg\{ \max\bigg\{\frac{\chi_{[0,T-\tau]}(t)\bigg(\lambda _{S_1}(t+\tau)S_1-\epsilon_{11}S_1\lambda_{V_1}(t+\tau)-\epsilon_{21}S_1\lambda_{F_1}(t+\tau)-x_1\bigg)}{2A_{1}},0 \bigg\}, \mu_{11}max\bigg\}\\
	\mu_{12}^{*} &= &\min\bigg\{ \max\bigg\{\frac{\chi_{[0,T-\tau]}(t)\bigg(\lambda _{S_2}(t+\tau)S_2-\epsilon_{12}S_1\lambda_{V_1}(t+\tau)-\epsilon_{22}S_2\lambda_{F_1}(t+\tau)-x_2\bigg)}{2A_{1}},0 \bigg\}, \mu_{12}max\bigg\}\\
	\mu_{2i}^{*} &= &\min\bigg\{ \max\bigg\{\frac{\chi_{[0,T-\tau]}(t)\bigg(\lambda _{S_i}(t+\tau)S_2-\gamma_{1i}S_1\lambda_{V_i}(t+\tau)-\gamma_{2i}S_2\lambda_{F_i}(t+\tau)-y_i\bigg)S_i}{2A_{2}},0 \bigg\}, \mu_{2i}max\bigg\}\\
		\mu_{3i}^{*} &= &\min\bigg\{ \max\bigg\{\frac{\chi_{[0,T-\tau_2]}(t)\bigg(\lambda _{I_i}(t+\tau_2)-\lambda _{R_i}(t+\tau_2)\bigg)I_i}{2A_{3}},0 \bigg\}, \mu_{3i}max\bigg\}\\
			\mu_{4i}^{*} &= &\min\bigg\{ \max\bigg\{\frac{\chi_{[0,T-\tau_3]}(t)\bigg(\lambda _{J_i}(t+\tau_3)-\lambda _{R_i}(t+\tau_3)\bigg)I_i}{2A_{4}},0 \bigg\}, \mu_{4i}max\bigg\}\\
	\end{eqnarray*}
	
	\section{Numerical Simulations}
	
		In this section, we perform numerical simulations to understand the age specific efficacies of vaccination and the treatment. This is done by studying the effect of control  on the  dynamics of the system. Let there exist a step size $ h > 0$ and $n > 0$ such that $T-t_0=nh$. Let $m= max(\tau,\tau_1, \tau_2, \tau_3)$. For programming point of view we consider m knots to left of $t_0$  and right of T and we obtain
the following partition:

$\Delta= \bigg(t_{-m}=-max(\tau,\tau_1,\tau_2,\tau_3).... < t_1 < t_0 =0 < t_1 ...<t_n = t_f(=T) < ....<t_{n+m} )\bigg)$.\\
Using combination of forward and backward
difference approximations,we simulate the results in matlab software. All the parameter values and the source from which they are taken is given in table 2. Initially, we work with the assumption that the efficacy of both the vaccine is 60 $\%$ and  later varying the  efficacy level of both the vaccines we plot the the changes in the infection and disease induced mortality. For the initial simulation we take the values of $A_i, i=1,2$,
the cost associated with vaccination as $10^2$. We also study the effects of optimal vaccination strategies on the dynamics of the disease under different vaccination coverages. In this context larger values of of the weights $A_i$ mean that
the cost associated with vaccination is expensive; hence, the vaccination coverages is less for
larger $A_i$. The values for weight constant associated with treatment for infected and hospitalized population $(A_i,i=3,4)$ are taken  as 200 and 100. The cost of treatment of the hospitalized population is taken lesser than that of treatment of infected population because it is assumed that all the facilities are available in the hospital. We have also assumed that the disease induced death rate of hospitalized is 100 times more than that of infected.

In simulation three control strategies are performed\\
\textbf{A:} Implementation of vaccination only strategy  to control the spread of COVID-19. \\
\textbf{B:} Implementation of treatment only strategy  to control the spread of COVID-19. \\
\textbf{C:} Implementation of both treatment and vaccination  strategies  to control the spread of COVID-19. \\

 \begin{table}[ht!]
     	\caption{}
     	\centering 
     	\begin{tabular}{|l|l|l|} 
     		\hline\hline
     		
     		\textbf{Parameters} &  \textbf{Value} &  \textbf{Source}\\  
     		\hline\hline 
     	
     		$\omega_i$ & 7.192 &\cite{samui2020mathematical} \\
     		\hline\hline
     		$\beta_{ij}$ &  (0.0175,0.0341,0.0319, 0.0339) & approximated from\cite{lee2012modeling} \\
     		\hline\hline
     		
     		$\mu$ & 0.062 &\cite{samui2020mathematical} \\
     		\hline\hline
     		$d_{11} $ &.000073 &\cite{chhetri2020optimal}  \\
     		\hline\hline
     		$d_{12}$ & 0.0000913    &\cite{chhetri2020optimal}\\
     		\hline\hline
     		$d_{21}$ & .0073 & assumed\\
     		\hline\hline
     		$d_{22}$& 0.00913& assumed\\
     			\hline\hline
     		$k$& 0.035&\cite{he2020seir} \\
     			\hline\hline
     		$\alpha_i$& (0.4, 0.5) &\cite{maki2020delayed} \\
     			\hline\hline
     		$\epsilon_{1i}, \epsilon_{2i}$& 0.2 & assumed  \\ 
     		\hline\hline
     		$\gamma_{1i}, \gamma_{2i}$& 0.2& assumed \\ 
     		\hline\hline
     		$\tau_1$ & 4 & \cite{maki2020delayed}\\
     		\hline\hline
     		$\tau_2$ & 12 & assumed\\
     		\hline\hline
     		$\tau_3$ & 12 & assumed\\
     		\hline\hline
     		$\gamma$& 0.07 & \cite{maki2020delayed} \\
     		\hline\hline
     		$\tau$ & 10 & \cite{elhia2013optimal}\\
     		\hline\hline
     		$A_i$ &$10^2  $ & assumed (baseline scenario) \\
     		\hline\hline
     		$A_3, A_4$ & 200, 100 & assumed\\ 
     		
     	\hline\hline
     		
     \end{tabular}
     \end{table} \vspace{.25cm}
     
\newpage
\subsection{Optimal control strategy}
In this section we  evaluate the role of each of the control strategy (vaccination and treatment)
 in reducing the COVID-19 burden for two specific age groups considered. Initially, we assume that the efficacy of both the vaccine is $60 \%$  and in later sections, we study the effect  of increasing the efficacy of vaccine on the infection and disease induced deaths. In figure 1 we plot the proportion of infected population with time for both the  age groups under different control strategies. In figure 2 and 3 the proportion of hospitalized and disease induced death curves are shown. From these figures we observe that the peak in the proportion of infected, hospitalized and deaths are minimum when  treatment and vaccination strategies are followed  together compared to the individual strategies alone. We also observe from figure 1 that with treatment only and combined strategy the peak of infection is reached faster in  time compared to no control and vaccination only strategy. The implementation of  optimal combined therapy leads to the reduction of approximately 50 $\%$ in the peak of infection for population of age between 0 to 40 followed by a  reduction of approximately 53 $\%$ for the second group ($> 40 )$ years compared to no control case. The reduction in the peaks of disease induced mortality for first and second age groups under the combined strategy are approximately  55 $\%$ and 62 $\%$ respectively compared to no control case.

 \newpage
 
 		 \begin{figure}[hbt!]
			\begin{center}
				\includegraphics[width=3.5in, height=3in, angle=0]{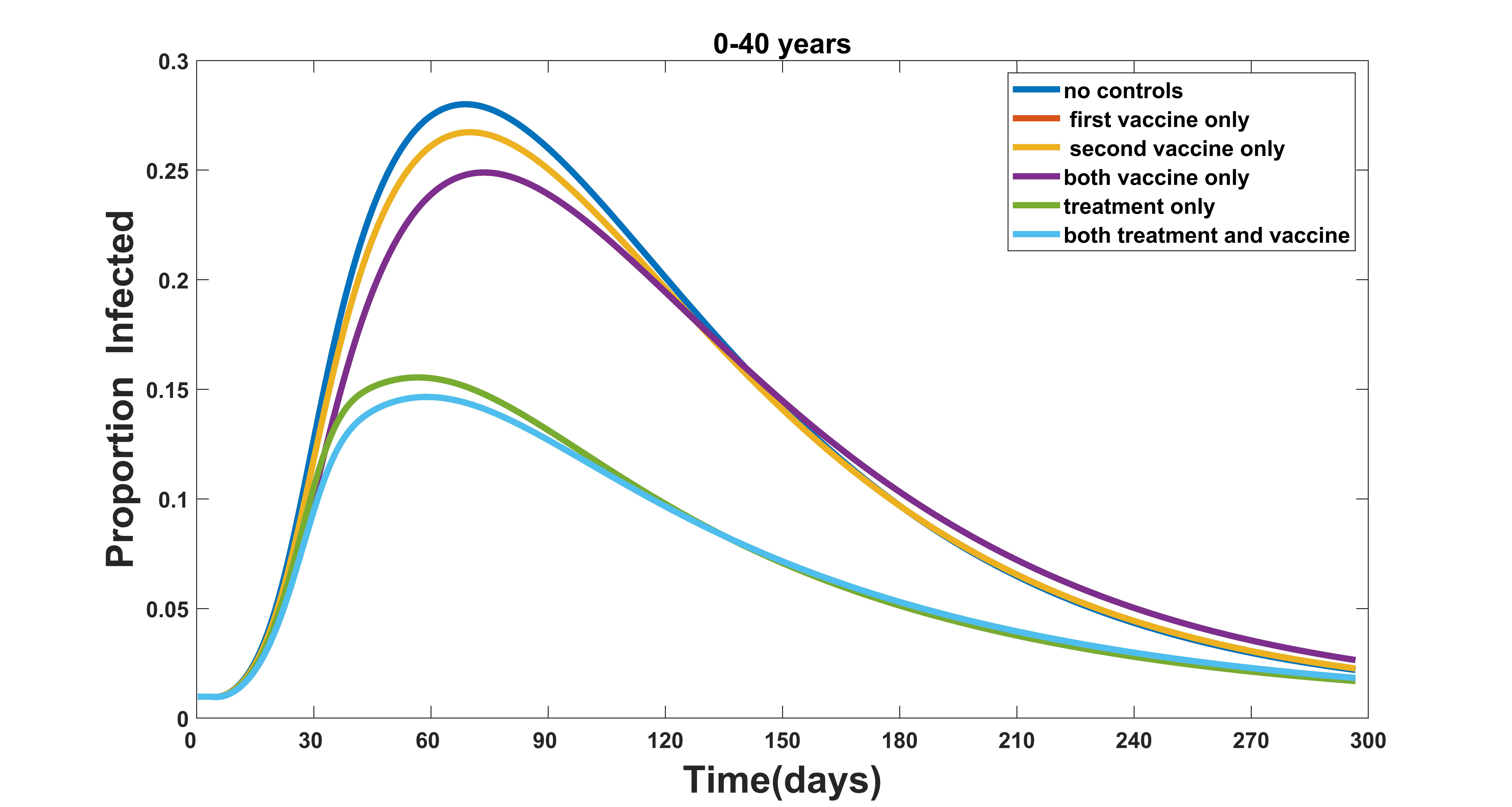}
				\hspace{-.4cm}
			\includegraphics[width=3.5in, height=3in, angle=0]{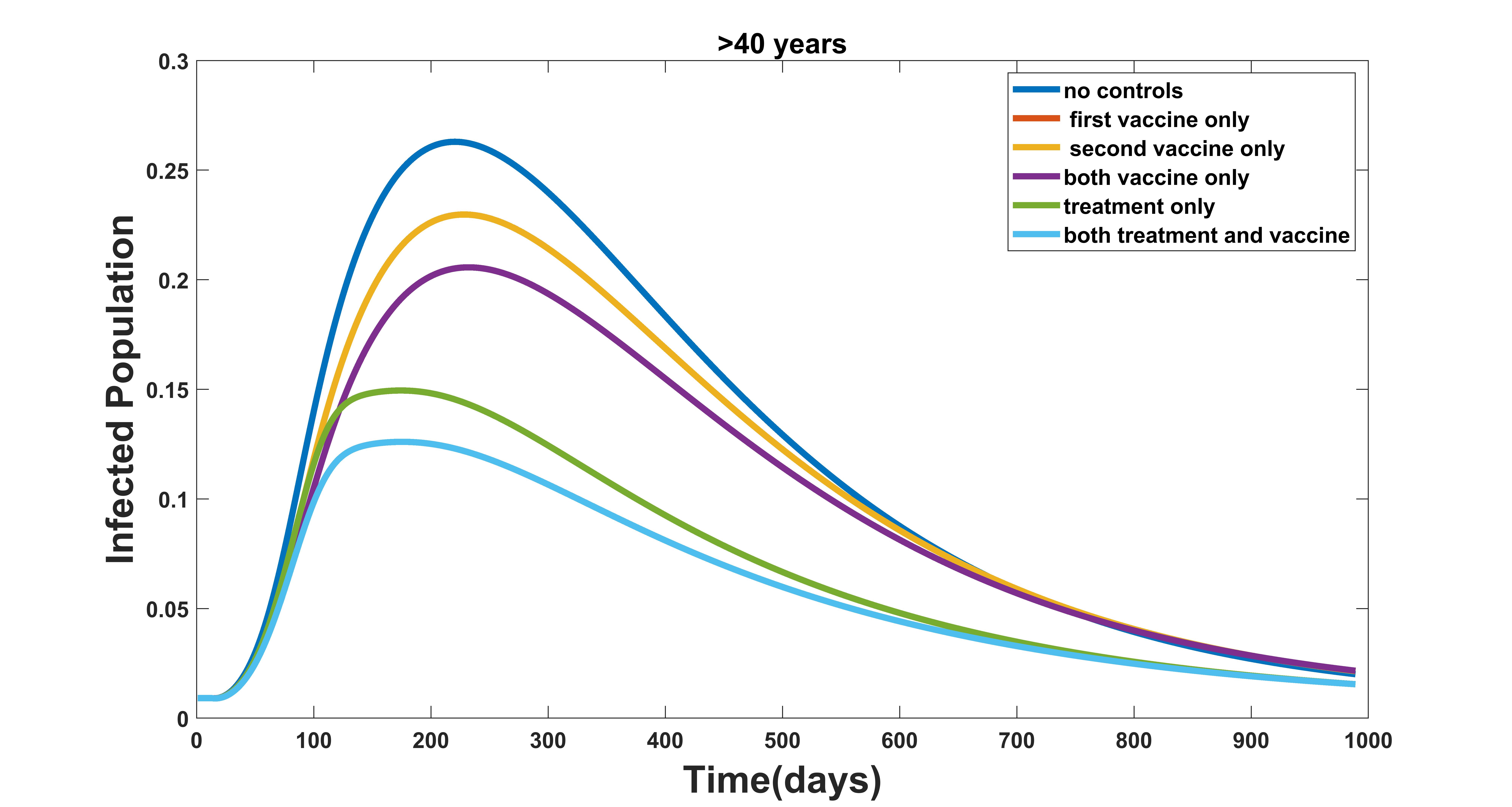}
			
			\caption{(a) Proportion of Infected population for first group \\
			(b)  Proportion of Infected population for second group}
				\label{b1}
			\end{center}
		\end{figure}

 		 \begin{figure}[hbt!]
			\begin{center}
				\includegraphics[width=3.5in, height=3in, angle=0]{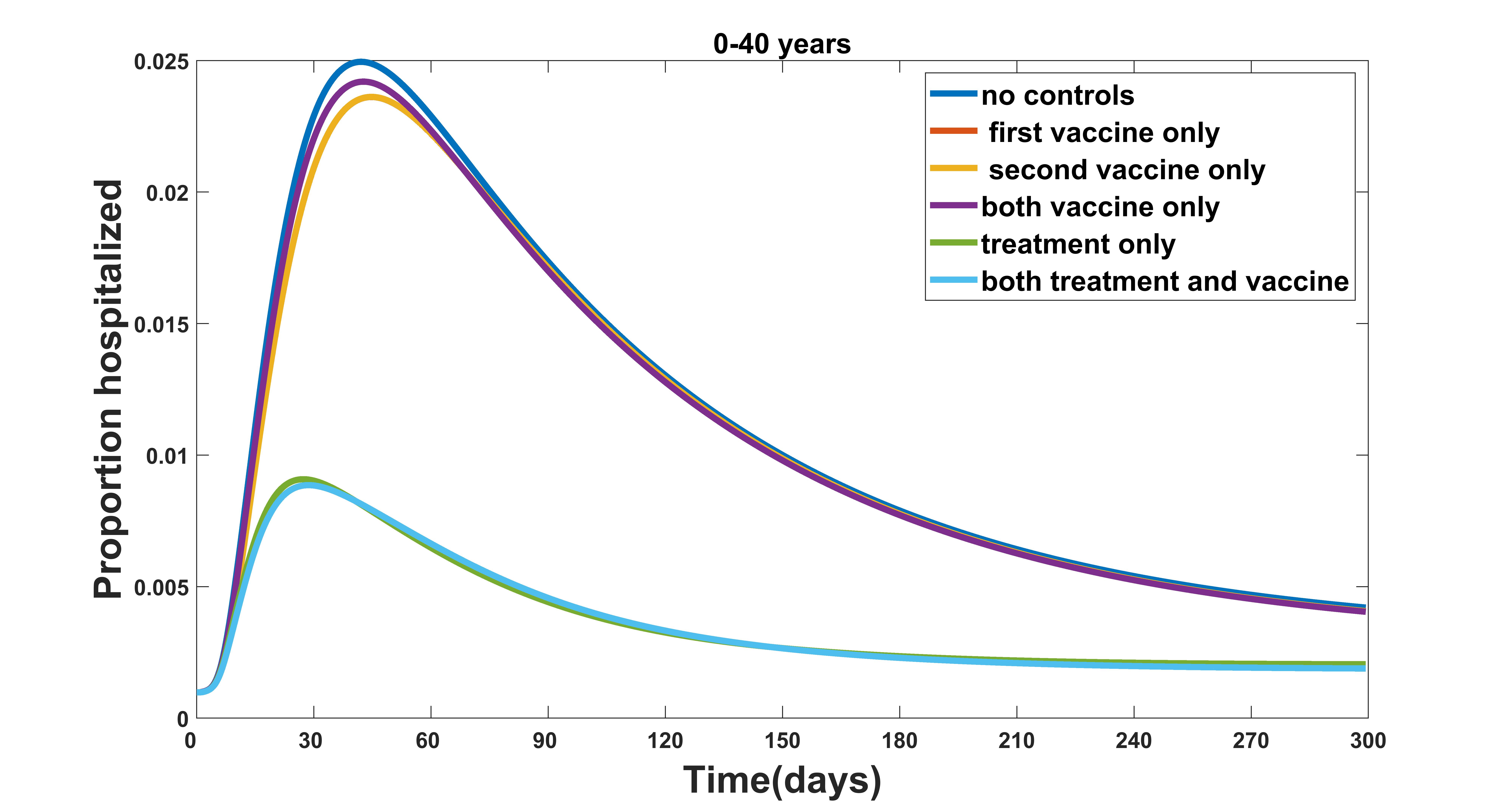}
				\hspace{-.4cm}
		    	\includegraphics[width=3.5in, height=3in, angle=0]{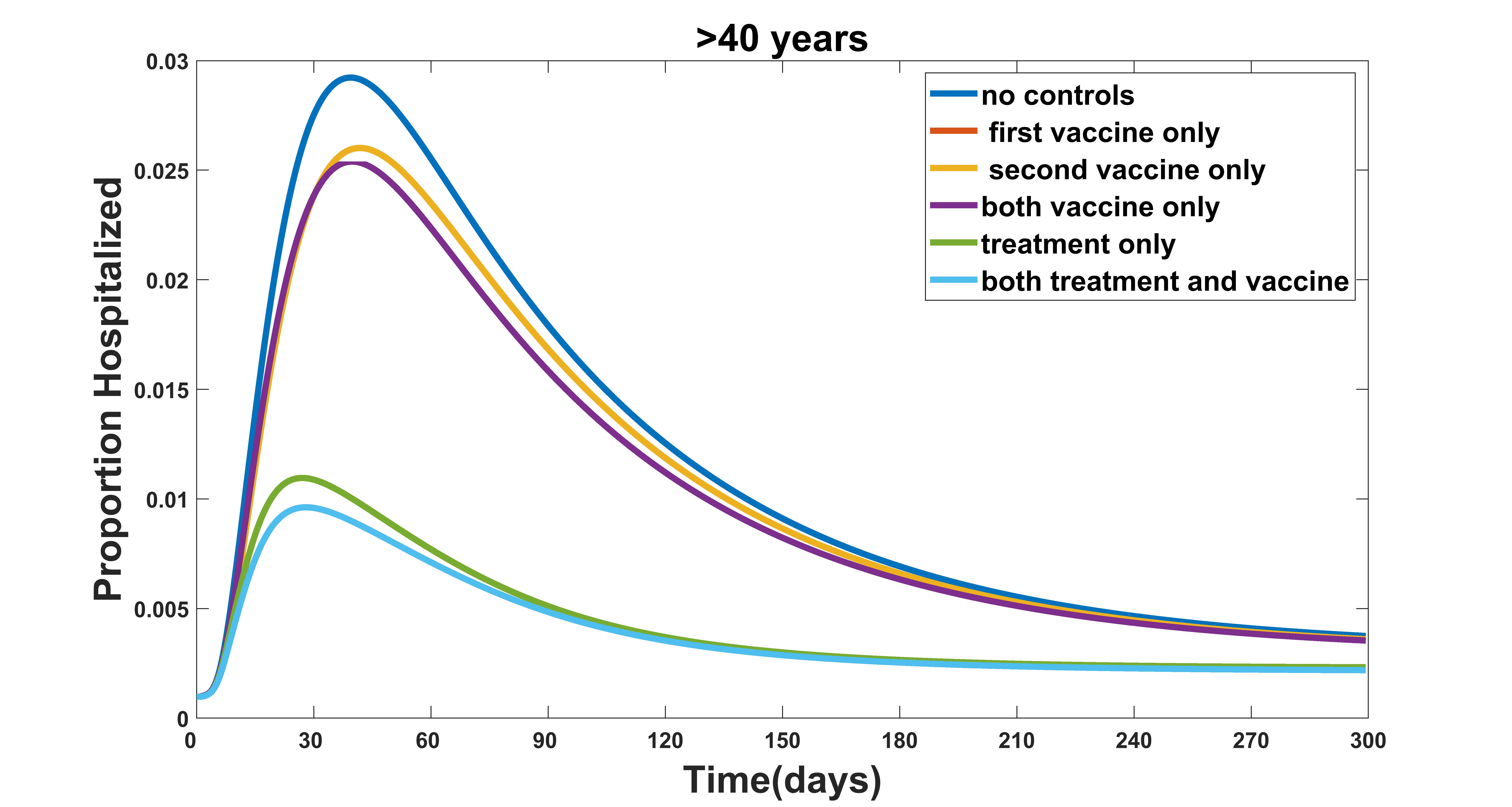}
				
				\caption{(a) Proportion of hospitalized population for first group \\
			(b)  Proportion of hospitalized population for second group}
				\label{b1}
			\end{center}
		\end{figure}

	 \begin{figure}[hbt!]
			\begin{center}
				\includegraphics[width=3.5in, height=3in, angle=0]{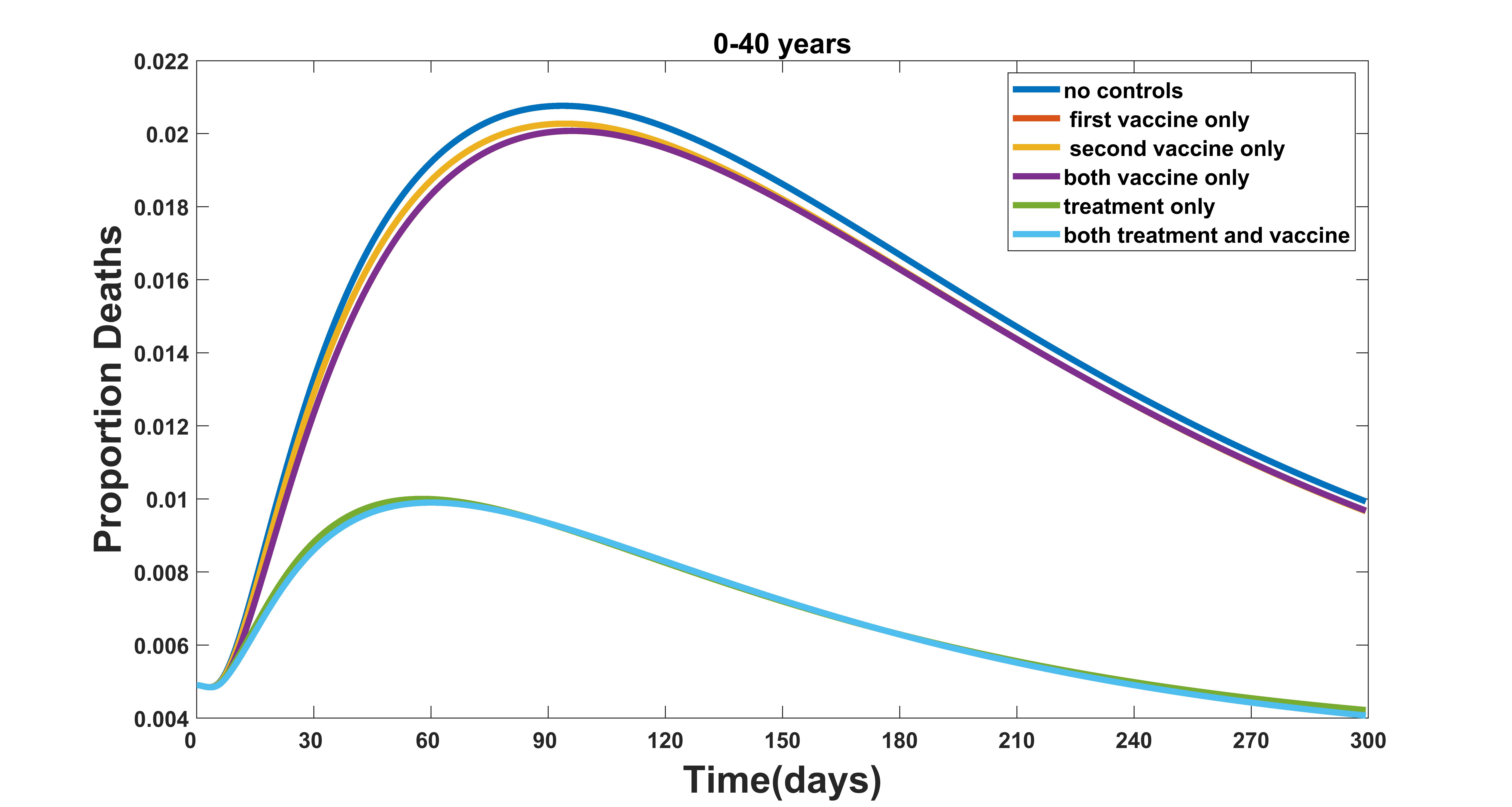}
				\hspace{-.4cm}
			\includegraphics[width=3.5in, height=3in, angle=0]{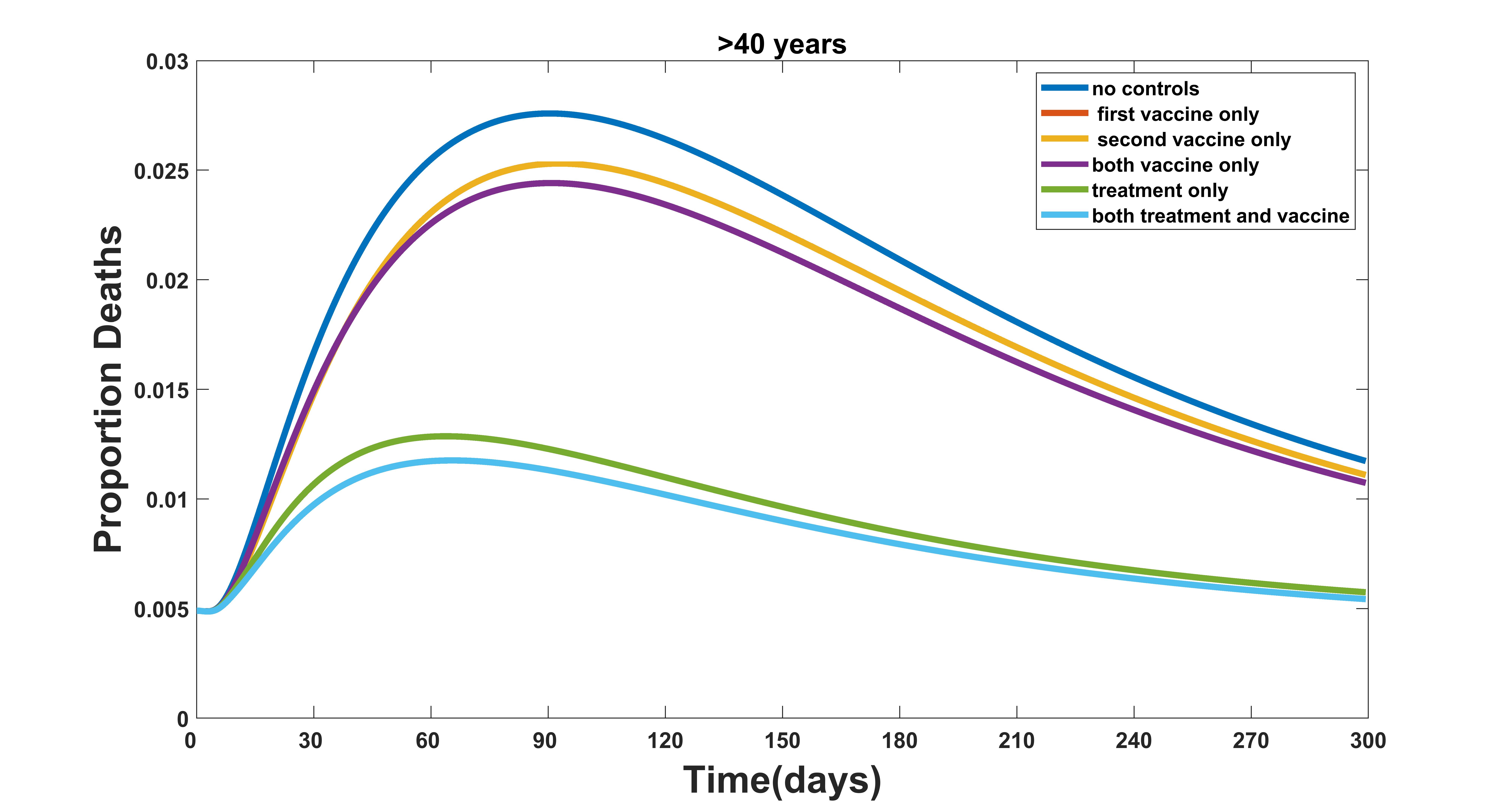}
				
				\caption{(a) Proportion of death population for first group \\
			(b)  Proportion of death population for second group}
				\label{b1}
			\end{center}
		\end{figure}
 
\newpage

 \newpage

Now we explore the role of age specific optimal combined strategies on the cumulative infection and disease induced mortality. In figure 4  we plot the cumulative infected and disease induced mortality considering optimal combined strategy. Comparing the cumulative infected population in absence of controls to  the cumulative infected population with optimal control strategies on the first age group, we observe from figure 4(a) that the reduction in the peaks of cumulative infection is approximately 21 percent. Similarly considering optimal control strategies on second age group,  we see that  there is approximately 25 percent reduction in the peaks of cumulative infection. We see that with optimal strategy reduction in the cumulative infection  is higher in case of second group.  Therefore, with this observation we claim that in order to reduce the infection  to maximum optimal control strategy should be  prioritized to the second age group. The cumulative disease induced mortality is plotted in figure 4(b) and the cumulative deaths decreased maximum when optimal combined strategy is prioritized to second group. 
 
 	 \begin{figure}[hbt!]
			\begin{center}
				\includegraphics[width=3.5in, height=3in, angle=0]{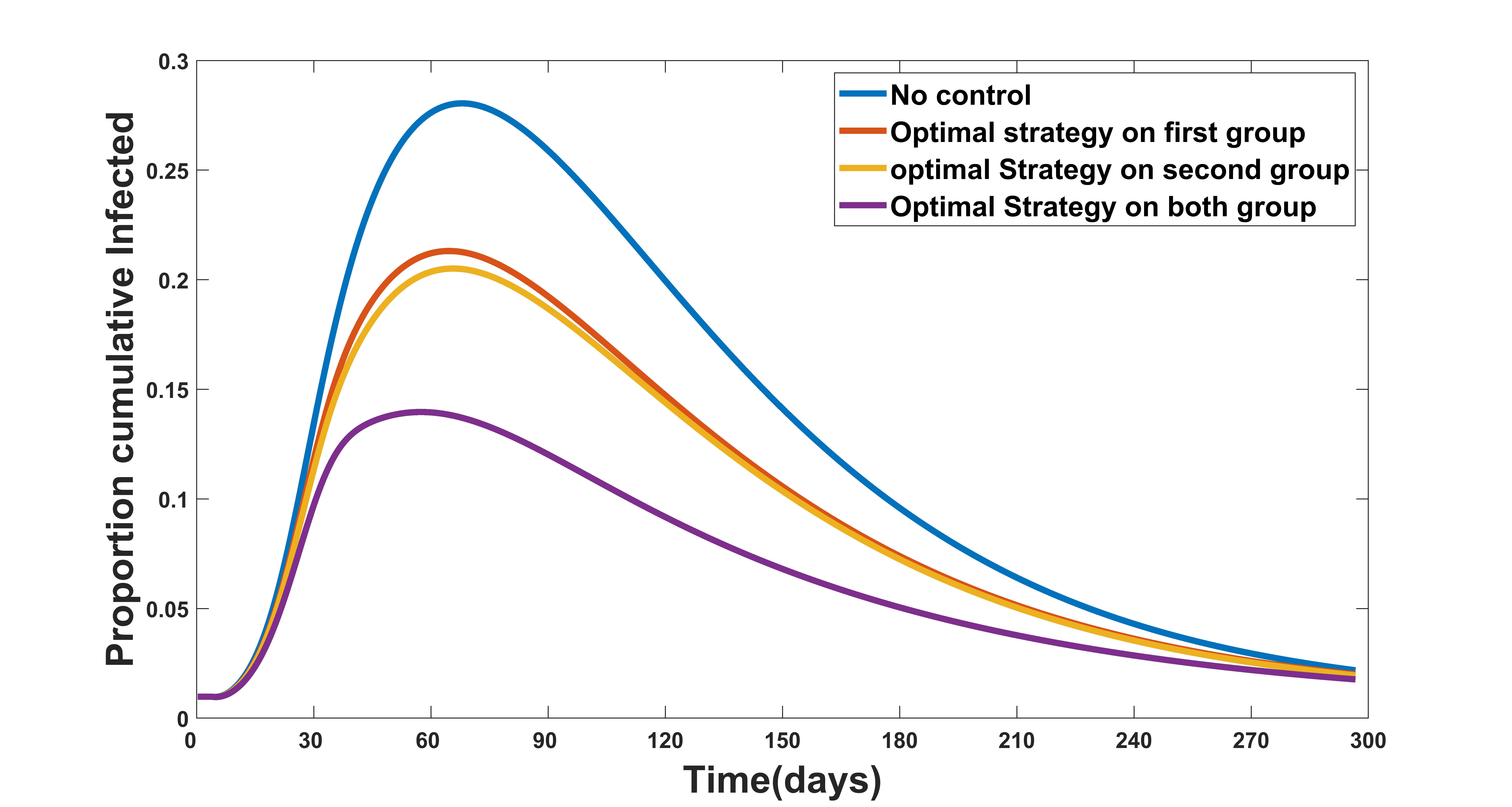}
				\hspace{-.4cm}
			\includegraphics[width=3.5in, height=3in, angle=0]{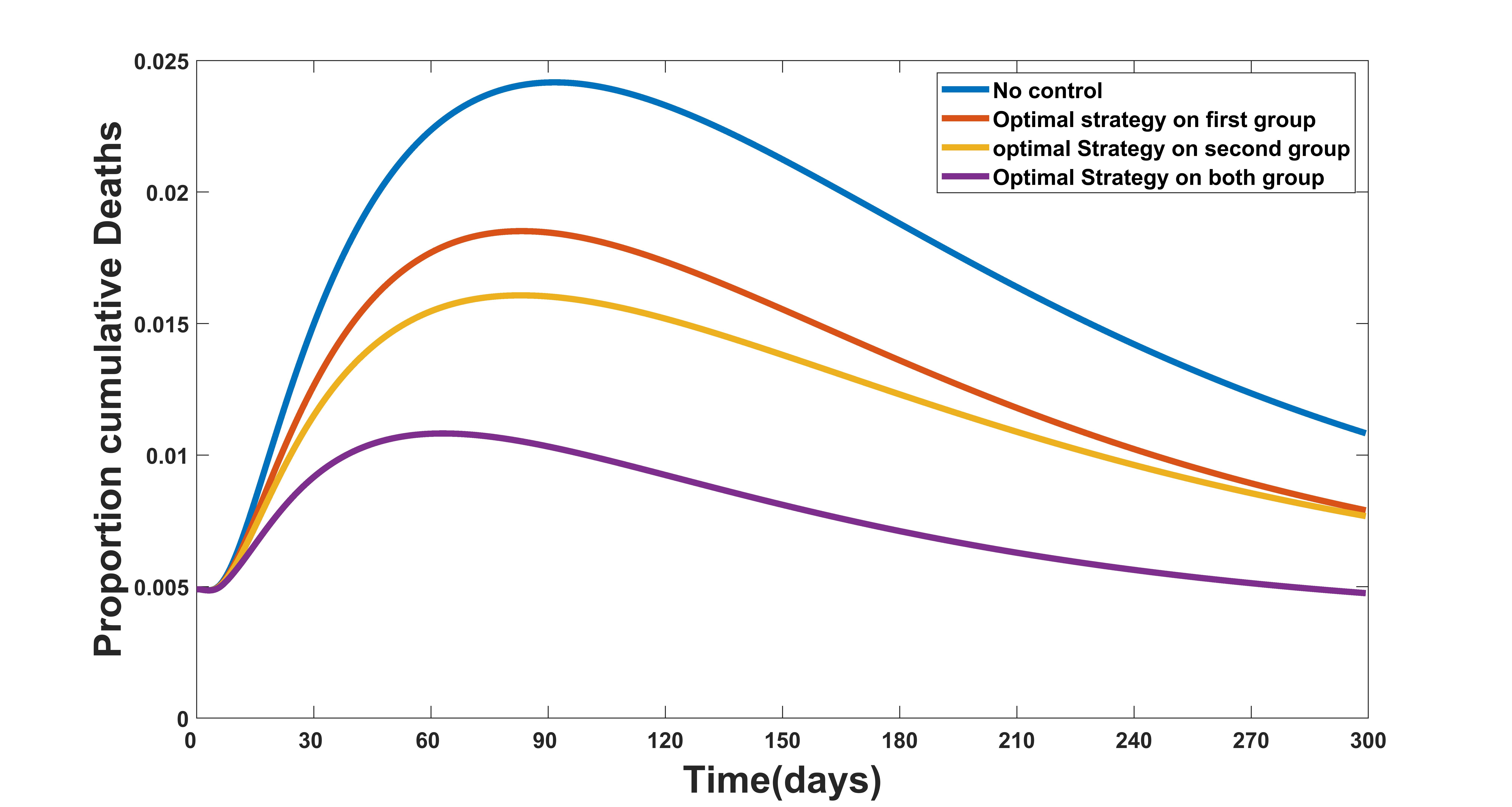}
				
				\caption{(a) Proportion of cumulative infection \\
			(b)  Proportion of cumulative deaths}
				\label{b1}
			\end{center}
		\end{figure}
 \newpage

 \subsection{Optimal Vaccination Strategies Under Different Vaccination Coverages}
 In the previous sections we had taken the baseline weight constant value related to vaccination ( $A_i$) as $10^2$ for i=1,2. In this section we study the effects of optimal vaccination strategy on the dynamics of the disease under different vaccination coverage. In the context larger values of of the weights $A_i$ means that
the cost associated with vaccination is expensive; hence, the vaccination coverages is less for
larger $A_i$. We assume that for the baseline value of the  weight constant the average vaccination coverages is about 60 $\%$ and as the cost of vaccination increases the average vacination coverage reduces.

In figure 5 we simulate the effect of varying the cost associated with  vaccination. As the value of weight constant increases, the cost of implementation of vaccination increases resulting in the reduction of vaccination rates. Due to this there is relatively higher number of infected population compared to the baseline case $(A_i=10^2)$. From figure 5 we see that the infection increase with the increase in the value of the cost for both the groups. There is almost 20 $\%$ and $5 \%$ increase in the infected population with the highest cost of vaccination for second and first age group respectively. The reason for the increase is that large coverages
of optimal age-specific vaccinations yield increased reductions in the overall
number of infected individuals.

\begin{figure}[hbt!]
			\begin{center}
				\includegraphics[width=3.5in, height=3in, angle=0]{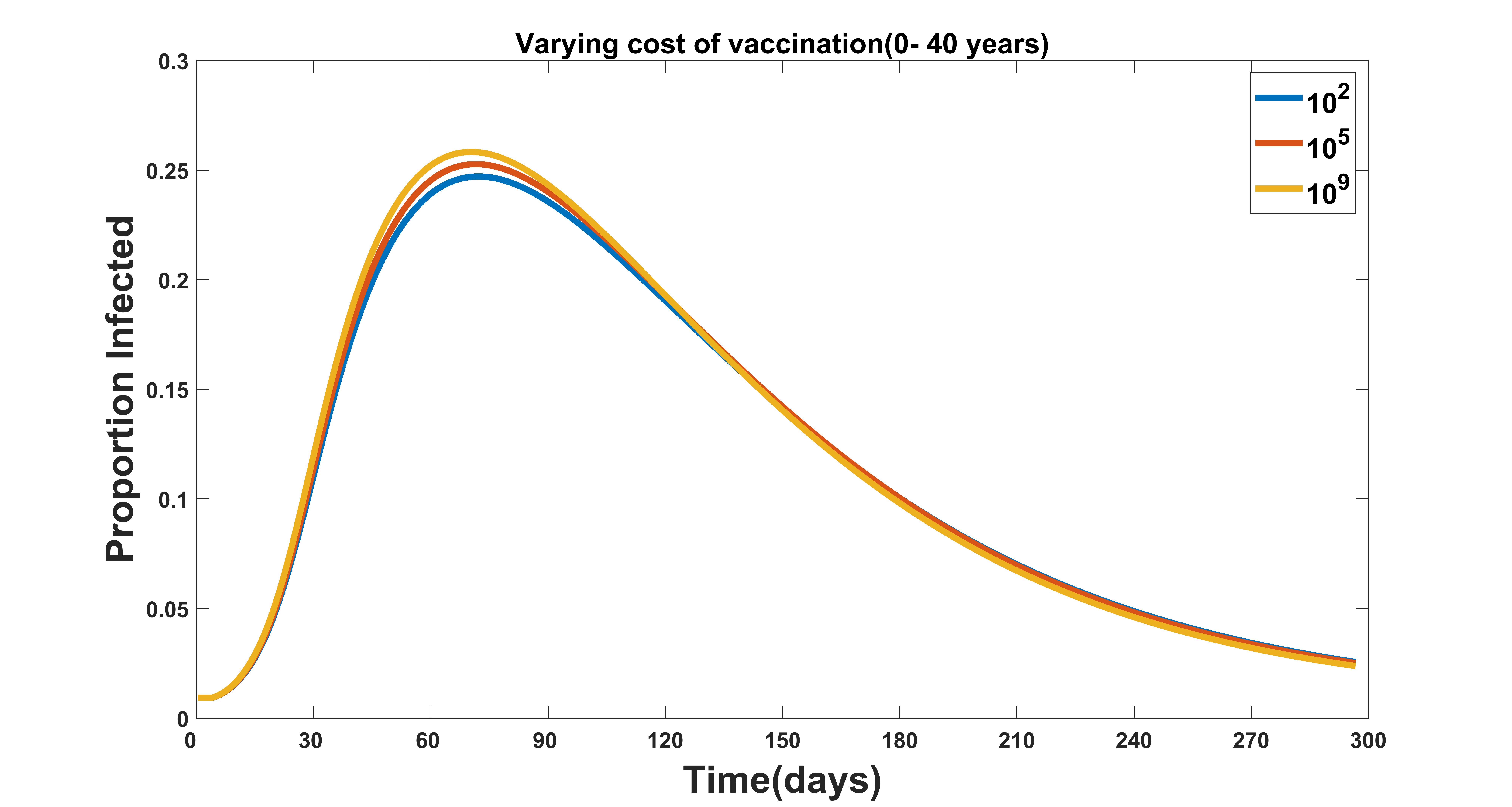}
				\hspace{-.4cm}
			\includegraphics[width=3.5in, height=3in, angle=0]{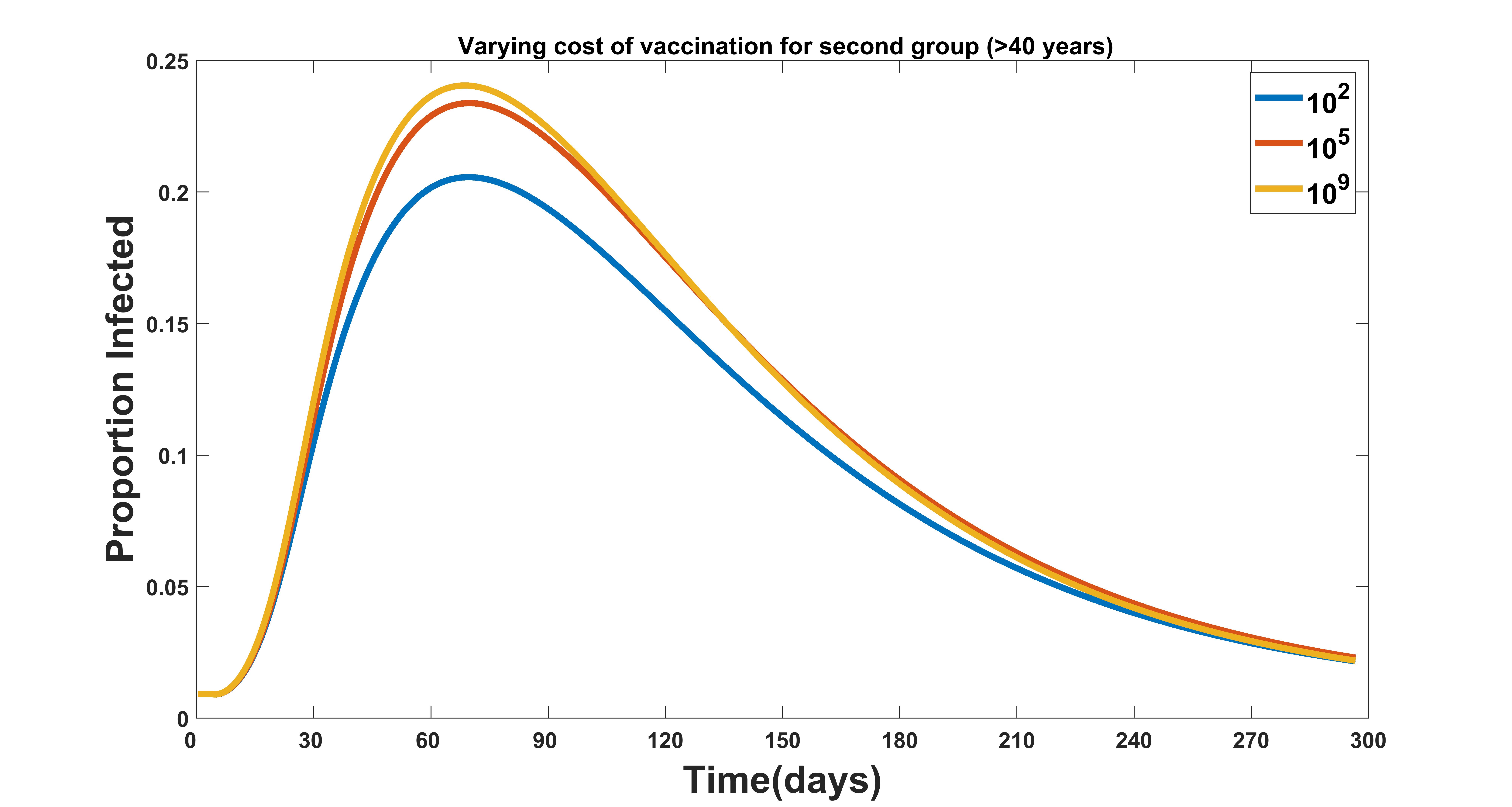}
				
				\caption{Proportion of infection varying weights \\
		}
				\label{b1}
			\end{center}
		\end{figure}
		
\subsection{Variation in Vaccination Efficacy}

Here we vary the efficacies of the vaccines and see the effects of varying efficacies in the proportion of infected and deaths. For the baseline scenario we assume  that the efficacy of both the vaccine is 60 $\%$ and then we vary the efficacies and see the relative changes in the proportion of infection and death with the baseline case. From figure 6 we see that as efficacy of vaccine increases, infection starts decreasing and it decreases the maximum with highest efficacy of the vaccine($90 \%$). Figure 7  shows that  disease induced mortality also decreases with increasing efficacy of the vaccine for both the age group considered.

\begin{figure}[hbt!]
			\begin{center}
				\includegraphics[width=3.5in, height=3in, angle=0]{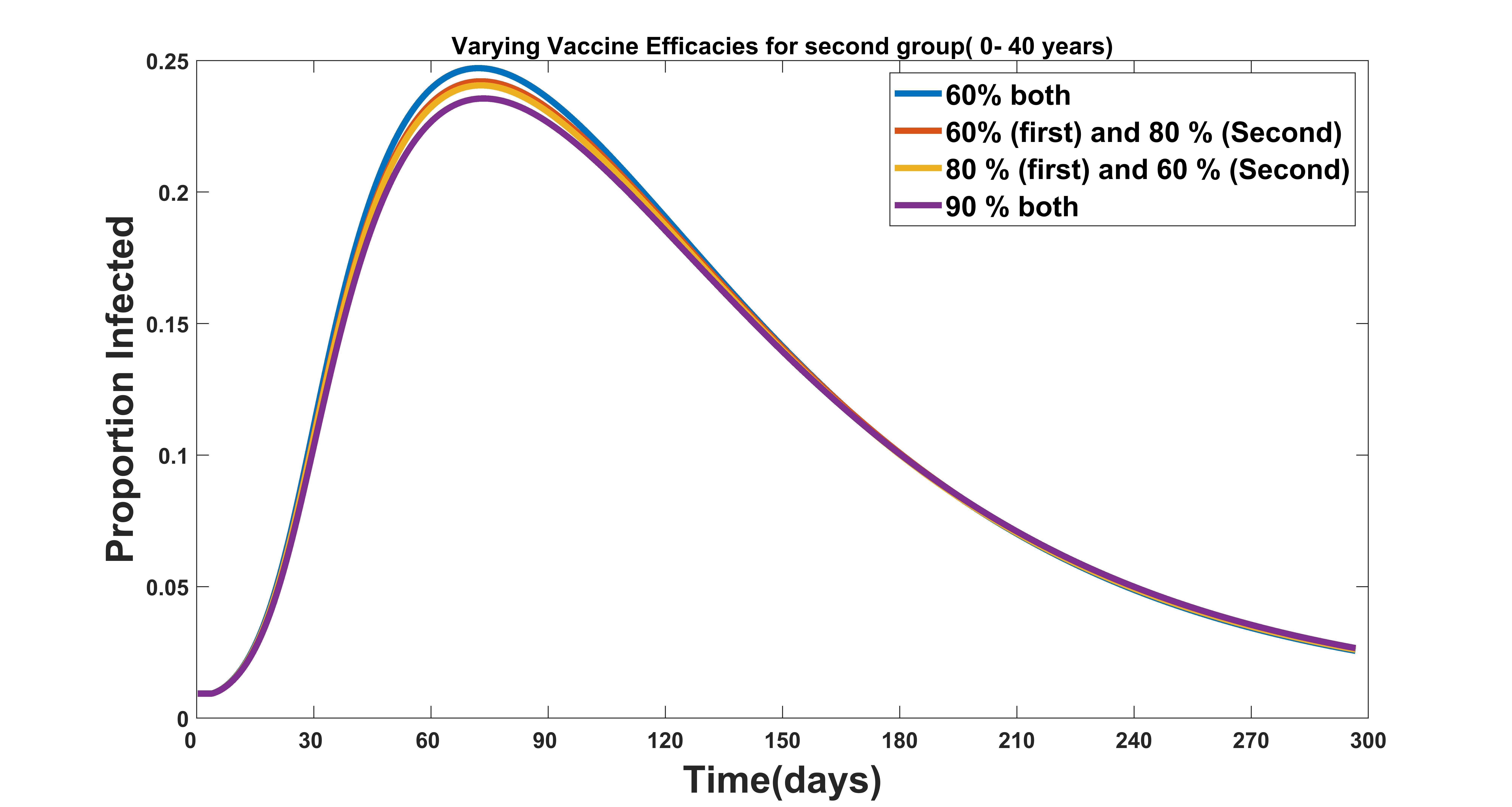}
				\hspace{-.4cm}
			\includegraphics[width=3.5in, height=3in, angle=0]{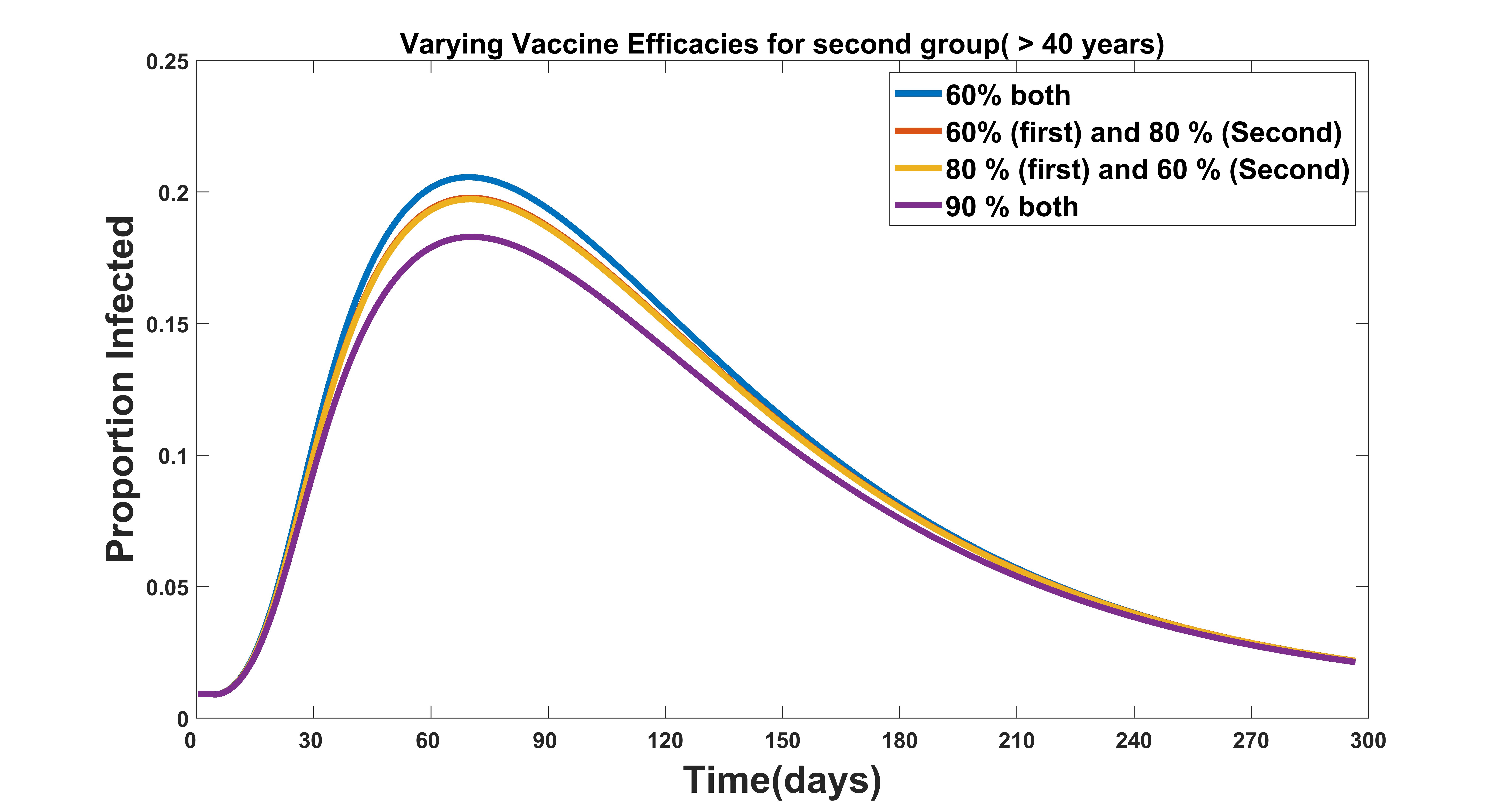}
				
				\caption{Proportion of infection varying vaccine efficacy \\
		}
				\label{b1}
			\end{center}
		\end{figure}

\begin{figure}[hbt!]
			\begin{center}
				\includegraphics[width=3.5in, height=3in, angle=0]{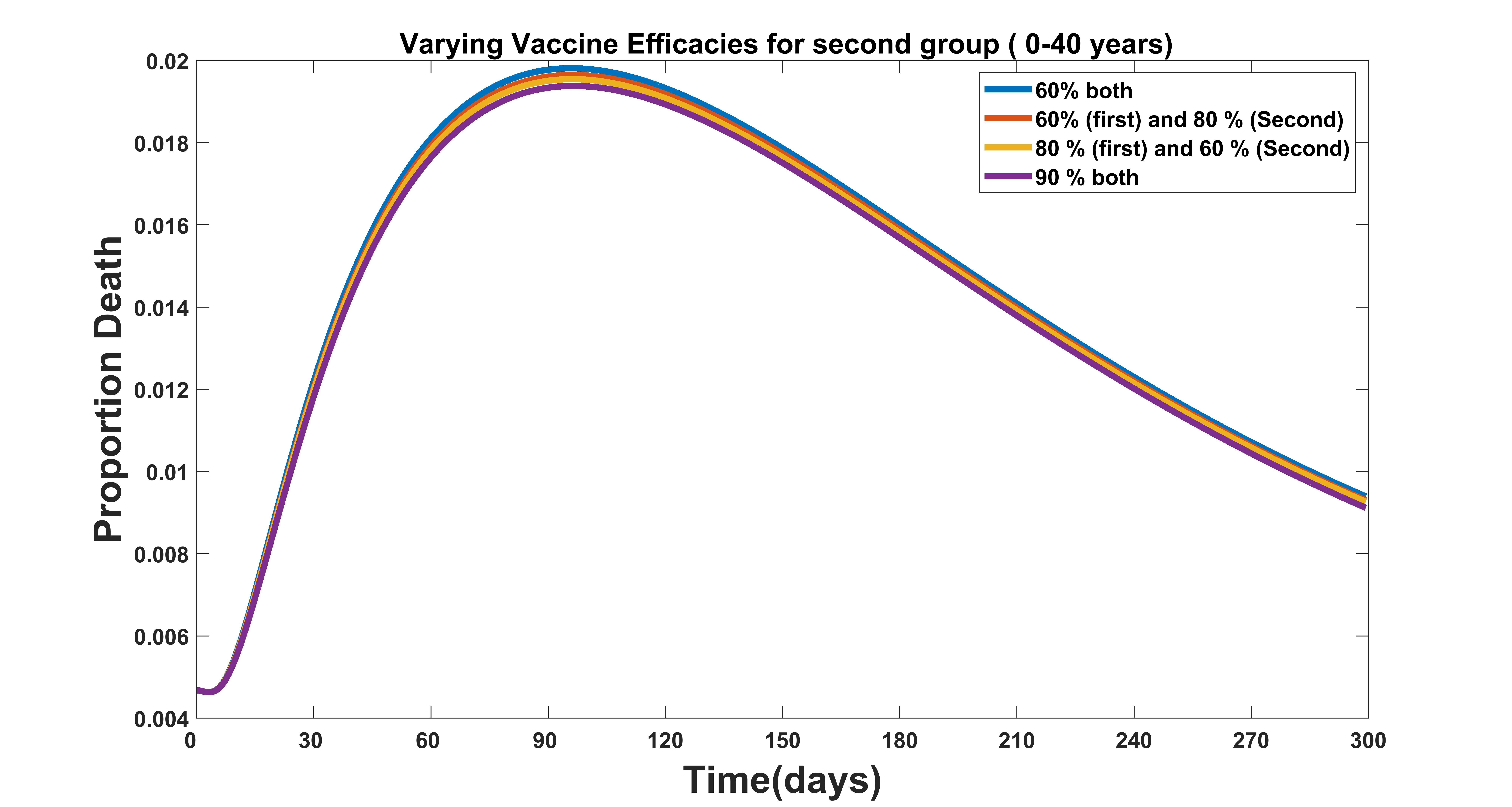}
				\hspace{-.4cm}
			\includegraphics[width=3.5in, height=3in, angle=0]{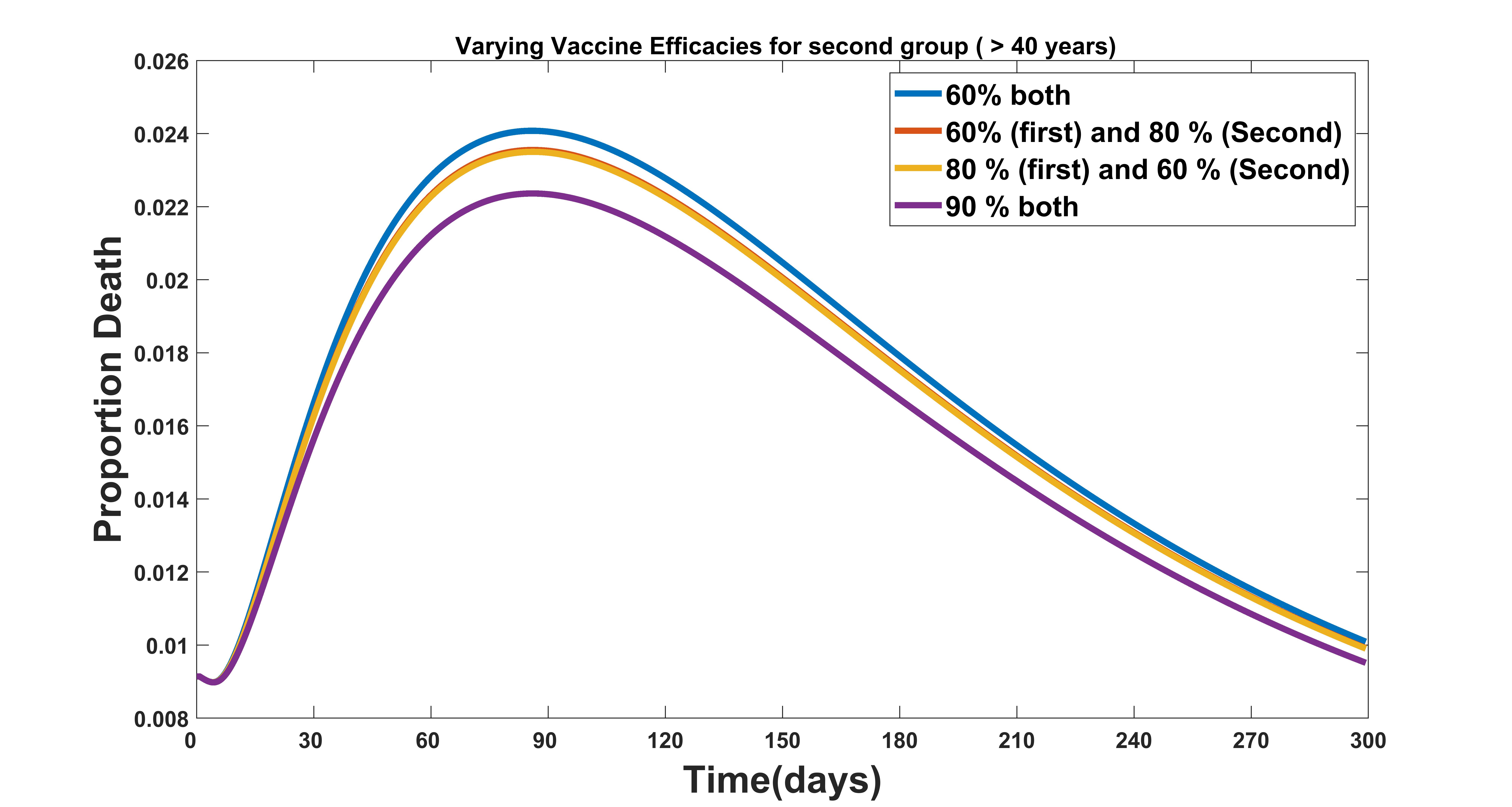}
				
				\caption{Proportion of deaths varying vaccine efficacy \\
		}
				\label{b1}
			\end{center}
		\end{figure}
		\newpage
		
		\subsection{The Effect of Optimal Age-Specific Vaccination Strategies Under Different Transmissibility
Levels ($R_0$) }
In this section we  study the dynamics of disease and the effect of vaccination strategy  with varying transmissibility $(R_0)$. Since the severity of the epidemic characterized by the high epidemic peaks which is measured by the higher values of $R_0$, therefore we will observe the prevalence of the cumulative
count of the disease  by varying the basic reproduction
number. From section 2.1 the basic reproduction number is given by,

 \begin{equation*}
         \mathbf{ R_{0}}= \frac{\beta_{11}k S_1^*}{(k+\mu)(d_{11}+\gamma+\alpha_1 e^{-\gamma\tau_1})} + \frac{\beta_{12}k S_2^*}{(k+\mu)(d_{12}+\gamma+\alpha_2 e^{-\gamma\tau_1})}
     \end{equation*}
     
   With varying values of $\mu=u=(0.062,0.1, 0.2)$ the values of $R_0$  were found to be (7.8, 4.5, 1.9) respectively. From figure 8  it can be observed that epidemic reaches it peak when R0 is around 2.5 with treatment only strategy. Whereas with vaccination only  strategy and combined optimal strategy  the peak is reached much faster. 
In figure 8 (a) we consider the efficacy of vaccine both the vaccine to be 60 $\%$ and in figure 8 (b) 90 $\%$.  varying $R_0$ in the x-axis in the between 0 to 10, we plot the proportion of cumulative infected  population considering different control strategies  for two age groups.  Our findings suggest that when the epidemic is mild $(R_0 \in (1, 1.5))$, all the control strategies works equally good. But as epidemic progresses the combined strategies(vaccination and treatment together) seems to work best in minimizing the cumulative infection. Comparing figure 8(a,b) it is observed that with increasing efficacy of the vaccine  the cumulative infection reduces.

\begin{figure}[hbt!]
			\begin{center}

					\includegraphics[width=3.5in, height=3in, angle=0]{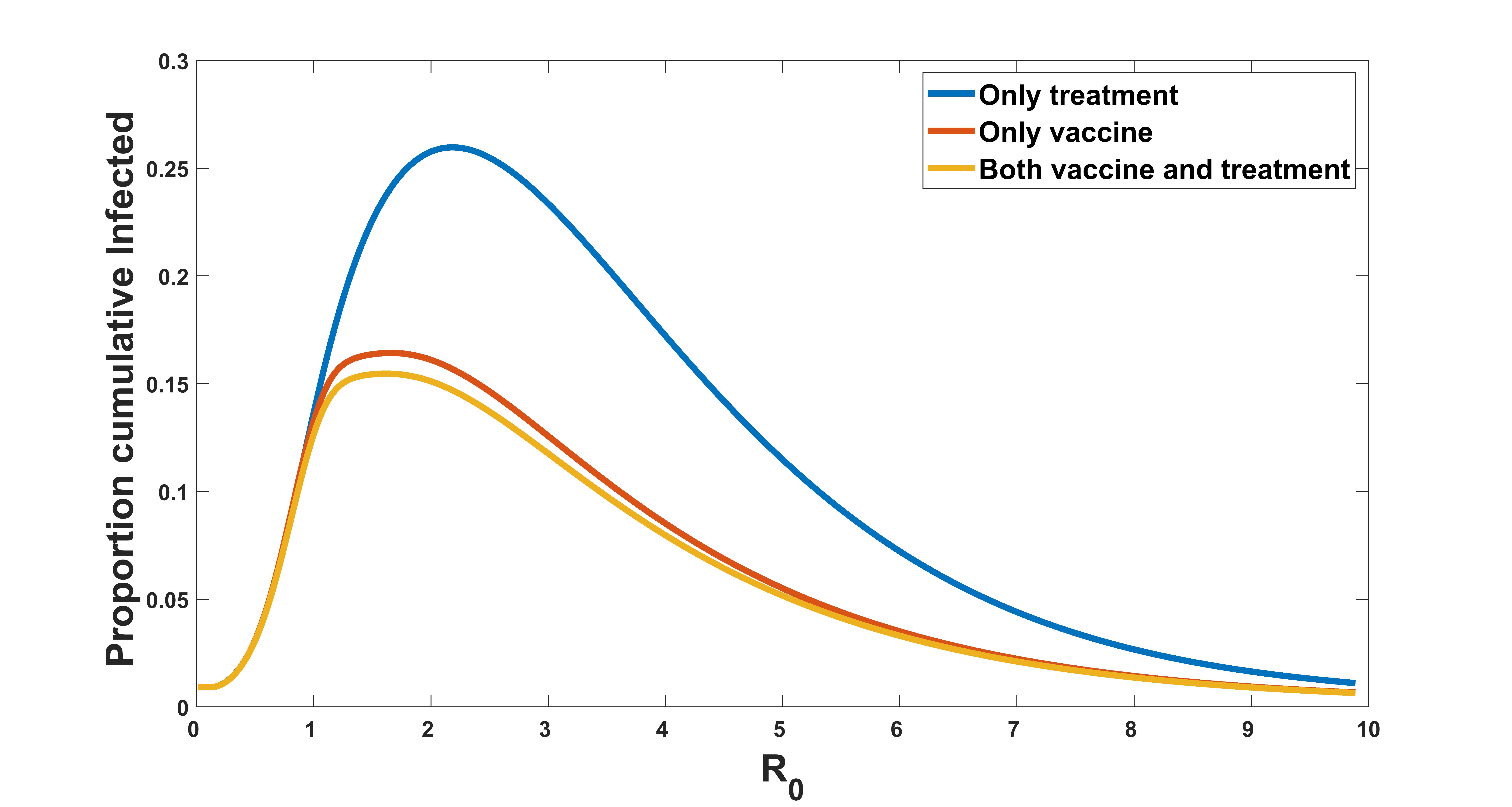}
				\hspace{-.4cm}
			\includegraphics[width=3.5in, height=3in, angle=0]{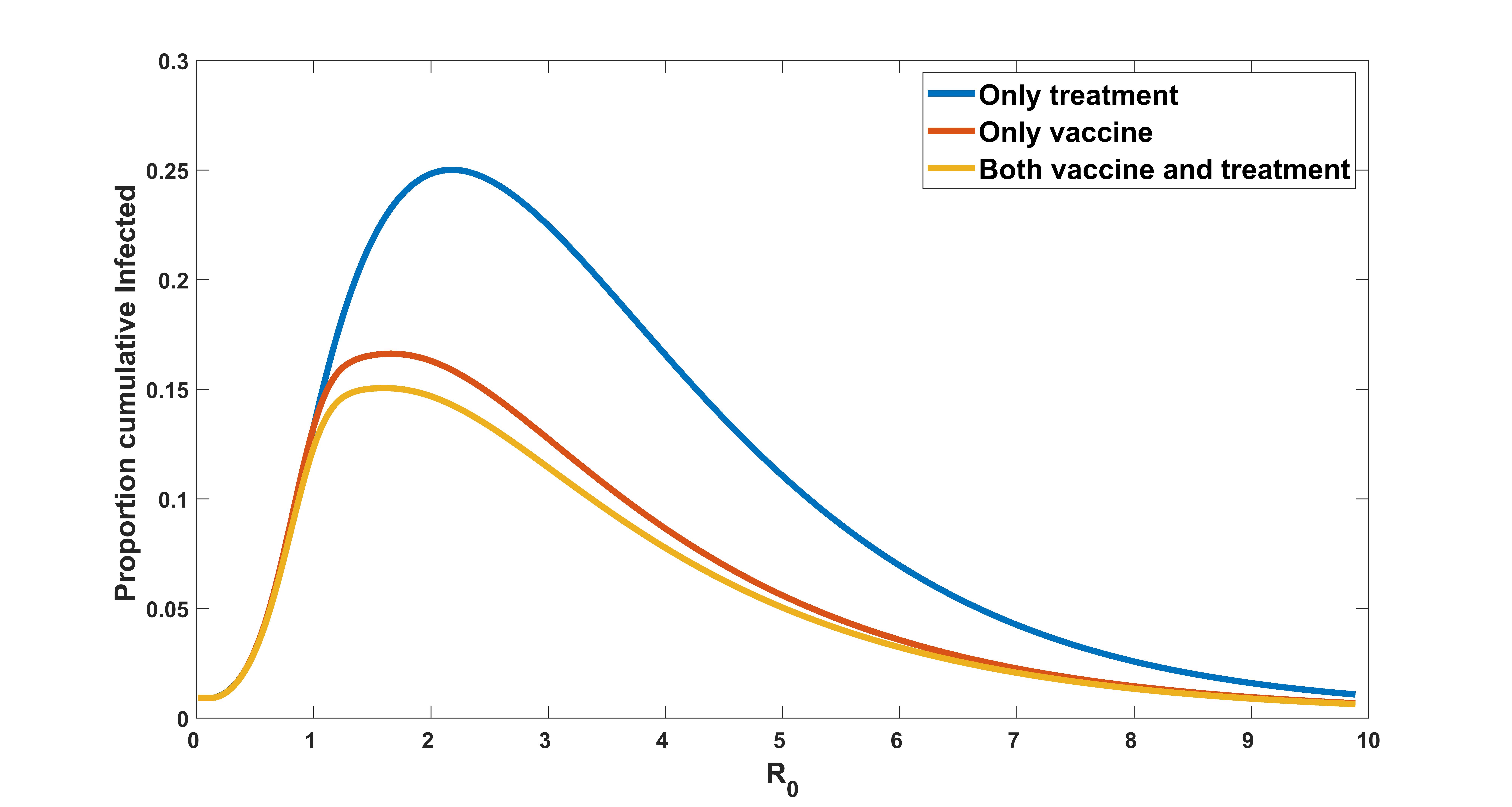}
			
			\caption{ Cumulative proportion of Infected population under different control strategies (a) 60 $\%$ vaccine efficacy (b) 90 $\%$ vaccine efficacy   \\
		}
				\label{b1}
			\end{center}
		\end{figure}

	\newpage
		\section{Discussion and Conclusion}
		In this work, the total population  was divided into 9 different compartments such as suceptibles$(S_i)$, vaccinated but not protected$(V_i)$, ineffectively vaccinated$(F_i)$, Protected$(P_i)$, exposed$(E_i)$, infected$(I_i)$, hospitalized$(J_i)$, recovered$(R_i)$ and deaths($D_i$) for i=1,2. Firstly,  an age specific model representing the dynamics of COVID-19 was  formulated and the positivity and boundedness of the model was established. Secondly to study the  effectiveness of the individual vaccine, combination vaccines and treatment an optimal  control problem with age specific transmission dynamics of COVID-19 was framed. After which numerical simulation are performed. In simulation three control strategies were performed\\
\textbf{A:} Implementation of vaccination only strategy  to control the spread of COVID-19. \\
\textbf{B:} Implementation of treatment only strategy  to control the spread of COVID-19. \\
\textbf{C:} Implementation of both treatment and vaccination  strategies  to control the spread of COVID-19. 

 The implementation of an age specific control strategies lead to the reduction  of infection, hospitalized population and disease induced deaths (figure 1,2,3). Compared to an individual vaccines strategy, combination vaccine strategy worked better in minimizing the infection and disease induced deaths. However, the  best  possible result in minimizing the peaks of infection and disease induced
deaths was achieved  when both vaccination and treatment strategies were used. This result is in similar lines to the results obtained in \cite{chhetri2020within,sari2017optimal}.

From figures 4,  it was observed that in order to reduce the cumulative infection and cumulative disease induced deaths to maximum optimal control strategy must be prioritized
to the second age group. When the cost of implementation of
vaccination increased there was
relatively higher number of infected population compared to the baseline case (figure 5).  The reason for these could be that with increasing cost  the vaccination coverage reduces as a result of which  there is increase in the number of infection. Increasing the efficacy of the vaccine also reduces the infection and disease induced deaths (figure 6,7).

 From figure 8  we observed that larger value of $R_0$ resulted in the larger pandemic sizes because of the rapid spread of the pandemic. When the epidemic was mild $R_0\in(1,1.5)$, all the control strategies worked equally good but as epidemic progressed over the time the best strategy to contain the size of epidemic was found to be the combined strategies(vaccination and treatment together).

		{\flushleft{  \textbf{ACKNOWLEDGEMENTS} }}\vspace{.25cm}

 The authors from SSSIHL  acknowledge the support of SSSIHL administration for this work.  \vspace{.25cm}

{\flushleft{  \textbf{DEDICATION} }}\vspace{.25cm}

The authors from SSSIHL and SSSHSS dedicate this paper to the founder chancellor of SSSIHL, Bhagawan Sri Sathya Sai Baba. The corresponding author also dedicates this paper to his loving elder brother D. A. C. Prakash who still lives in his heart and the first author dedicates this paper to his loving Grandmother.

\bibliographystyle{amsplain}
\bibliography{reference}
     
     \end{document}